%% file: main.tex
\documentclass[journal]{vgtc}                     
\usepackage{indentfirst}

\onlineid{2445}

\vgtccategory{Research}

\vgtcpapertype{technique}

\title{Explorable INR: An Implicit Neural Representation for Ensemble Simulation Enabling Efficient Spatial and Parameter Exploration}

\author{%
    Yi-Tang Chen,
    Haoyu Li,
    Neng Shi,
    Xihaier Luo,
    Wei Xu,
    Han-Wei Shen
}

\authorfooter{
  \item
  	Yi-Tang Chen is with The Ohio State University.
  	E-mail: chen.10522@buckeyemail.osu.edu
}

\abstract{%
With the growing computational power available for high-resolution ensemble simulations in scientific fields such as cosmology and oceanology, storage and computational demands present significant challenges. Current surrogate models fall short in the flexibility of point- or region-based predictions as the entire field reconstruction is required for each parameter setting, hence hindering the efficiency of parameter space exploration. Limitations exist in capturing physical attribute distributions and pinpointing optimal parameter configurations. In this work, we propose \textit{Explorable INR}, a novel implicit neural representation-based surrogate model, designed to facilitate exploration and allow point-based spatial queries without computing full-scale field data. In addition, to further address computational bottlenecks of spatial exploration, we utilize probabilistic affine forms (PAFs) for uncertainty propagation through Explorable INR to obtain statistical summaries, facilitating various ensemble analysis and visualization tasks that are expensive with existing models. Furthermore, we reformulate the parameter exploration problem as optimization tasks using gradient descent and KL divergence minimization that ensures scalability. We demonstrate that the Explorable INR with the proposed approach for spatial and parameter exploration can significantly reduce computation and memory costs while providing effective ensemble analysis.
}


\graphicspath{{figs/}{figures/}{pictures/}{images/}{./}} 

\usepackage{tabu}                      
\usepackage{booktabs}                  
\usepackage{lipsum}                    
\usepackage{mwe}                       
\usepackage{amsfonts} 
\usepackage{amsmath}
\usepackage{multirow}
\usepackage{graphicx}
\usepackage{siunitx}

\usepackage{mathptmx}                  

\begin{document}


\firstsection{Introduction}
\label{sect:intro}
\maketitle

\input{tex/1_introduction}
\input{tex/2_related_work}
\input{tex/3_method}
\input{tex/4_results}
\input{tex/5_conclusion}

\acknowledgments{
This work is supported in part by the US Department of Energy SciDAC program DE-SC0021360 and DE-SC0023193, National Science Foundation Division of Information and Intelligent Systems IIS-1955764, and Los Alamos National Laboratory Contract C3435. Brookhaven National Laboratory is supported by the DOE Office of Science under Contract No. DE-SC0012704 and FWP No. CC122.
}

\bibliographystyle{abbrv-doi-hyperref}

\bibliography{main}

\end{document}


\maketitle


\section{Mathematical Proof of Remark 1}
From a theoretical perspective, our hybrid approach, which combines feature grids and planes, is expected to surpass the K-planes model in terms of performance.
This is due to the mathematical properties of the interpolated feature vector in our method, which forms a polynomial of a higher degree compared to that derived from the K-plane, enabling the capture of more intricate details from the training data.
In the feature grid-based models, the interpolation function of a 3D feature grid can be represented as $f_{xyz}(x,y,z)$, where $(x, y, z)$ is the spatial coordinate, with $f_{xyz}(x,y,z)$ being a cubic polynomial with the highest total degree term being $xyz$.
The interpolation functions for 2D feature planes, namely $f_{xy}(x,y)$, $f_{yz}(y,z)$, and $f_{xz}(x,z)$, are quadratic polynomials with the highest total degree terms being $xy$, $yz$, and $xz$, respectively.
Through the Hadamard Product, the fused spatial feature vector $F_{sp}(x,y,z)$ for our hybrid grid and planes is given by
\begin{equation}
    F_{sp}(x,y,z) = f_{xyz}(x,y,z) \odot f_{xy}(x,y) \odot f_{yz}(y,z) \odot f_{xz}(x,z),
\end{equation}
which results in a polynomial of degree 9, with the highest total degree term $x^{3}y^{3}z^{3}$ for each element in the feature vector.
In the feature decoder, the polynomials are linearly combined and passed through an activation function, neither of which changes the polynomial degree in most cases.
Thus, the output still preserves the learning capability of a polynomial of degree 9. 
Concerns about high-degree polynomials leading to overfitting in uniform regions are mitigated by the feature grid's ability to learn features that can cancel out the terms with the highest degree in the polynomial, enabling our model to use lower-degree polynomials to describe homogeneous regions effectively.
On the other hand, the K-plane model utilizes multiresolution 2D planes, where the fused feature vector $F_{plane}^{i}(x,y,z)$ for each resolution $i$ is a polynomial of degree 6.
Consequently, each element in the concatenated feature vector remains a polynomial of degree 6, suggesting a potentially lower capacity to learn complex details compared to our proposed method.

\begin{table}[htbp]
\centering
\caption{The impact of the length of feature vectors. SDIM stands for spatial feature vector dimension; PDIM stands for parameter feature vector dimension. The experiments are also conducted on Nyx and MPAS-Ocean, and the comparison metrics are PSNR and MD.}
\label{tab:spdim_exp}
\resizebox{\columnwidth}{!}{%
\begin{tabular}{c|c|c|ccc}
\multicolumn{1}{l|}{\multirow{2}{*}{SDIM}} & \multirow{2}{*}{Dataset}    & \multirow{2}{*}{Metrics} & \multicolumn{3}{c}{PDIM}                                                    \\ \cline{4-6} 
\multicolumn{1}{l|}{}                      &                             &                         & \multicolumn{1}{c|}{16}              & \multicolumn{1}{c|}{32}     & 64     \\ \hline
\multirow{4}{*}{16}                        & \multirow{2}{*}{Nyx}        & PSNR (dB)               & \multicolumn{1}{c|}{41.86}           & \multicolumn{1}{c|}{42.02}  & 40.08  \\ \cline{3-6} 
                                           &                             & MD                      & \multicolumn{1}{c|}{0.1398}          & \multicolumn{1}{c|}{0.1359} & 0.1401 \\ \cline{2-6} 
                                           & \multirow{2}{*}{MPAS-Ocean} & PSNR (dB)               & \multicolumn{1}{c|}{48.80}           & \multicolumn{1}{c|}{49.37}  & 49.07  \\ \cline{3-6} 
                                           &                             & MD                      & \multicolumn{1}{c|}{0.1890}          & \multicolumn{1}{c|}{0.1953} & 0.1801 \\ \hline
\multirow{4}{*}{32}                        & \multirow{2}{*}{Nyx}        & PSNR (dB)               & \multicolumn{1}{c|}{42.52}           & \multicolumn{1}{c|}{43.45}  & 43.12  \\ \cline{3-6} 
                                           &                             & MD                      & \multicolumn{1}{c|}{0.1372}          & \multicolumn{1}{c|}{0.1084} & 0.1143 \\ \cline{2-6} 
                                           & \multirow{2}{*}{MPAS-Ocean} & PSNR (dB)               & \multicolumn{1}{c|}{50.28}           & \multicolumn{1}{c|}{49.85}  & 50.02  \\ \cline{3-6} 
                                           &                             & MD                      & \multicolumn{1}{c|}{0.1827}          & \multicolumn{1}{c|}{0.1786} & 0.1877 \\ \hline
\multirow{4}{*}{64}                        & \multirow{2}{*}{Nyx}        & PSNR (dB)               & \multicolumn{1}{c|}{\textbf{45.31}}  & \multicolumn{1}{c|}{43.82}  & 43.14  \\ \cline{3-6} 
                                           &                             & MD                      & \multicolumn{1}{c|}{\textbf{0.1015}} & \multicolumn{1}{c|}{0.1230} & 0.1133 \\ \cline{2-6} 
                                           & \multirow{2}{*}{MPAS-Ocean} & PSNR (dB)               & \multicolumn{1}{c|}{\textbf{50.81}}  & \multicolumn{1}{c|}{50.74}  & 50.49  \\ \cline{3-6} 
                                           &                             & MD                      & \multicolumn{1}{c|}{\textbf{0.1721}} & \multicolumn{1}{c|}{0.1779} & 0.1726
\end{tabular}%
}
\end{table}

\section{Feature Vector Dimension Exploration}
In Explorable INR, spatial coordinates and simulation parameters undergo spatial and parameter encoding, respectively, to generate spatial and parameter feature vectors. 
The dimensions of these feature vectors are hyperparameters of Explorable INR, necessitating an investigation into their impact on model performance. 
To explore the influence of feature vector dimensions, we examined sizes of 16, 32, and 64 for both spatial and parameter feature vectors. 
We utilized the Nyx and MPAS-Ocean datasets, employing data-level metrics such as PSNR and MD to evaluate the performance of various feature vector size combinations. 
As demonstrated in \cref{tab:spdim_exp}, optimal performance was achieved with a spatial feature vector size of 64 and a parameter feature vector size of 16. 
The table can be interpreted from two perspectives. 
First, when using a constant parameter feature vector size, increasing the spatial feature vector size correlated with improved performance. 
Second, when maintaining a constant spatial feature vector size, optimal performance was observed with parameter feature vector sizes of 16 or 32. 
A possible explanation for this phenomenon is that, due to limited samples in the parameter domain, the parameter feature vector can only be trained on these sparse samples. However, a longer feature vector may require more data to learn the correct distribution accurately. 
Thus, a long parameter feature vector may potentially impede model performance.

\begin{table}[htbp]
\centering
\caption{The proposed strategy utilizes the 4 strategies from the main text. S1-S4 are the ablation study on the 4 strategies. The results are evaluated on the Nyx and MPAS-Ocean datasets with metrics PSNR and MD.}
\label{tab:ablation_study}
\resizebox{\columnwidth}{!}{%
\begin{tabular}{c|cc|cc}
                            & \multicolumn{2}{c|}{Nyx}                              & \multicolumn{2}{c}{MPAS-Ocean}                        \\ \cline{2-5} 
Strategy                    & \multicolumn{1}{c|}{PSNR}           & MD              & \multicolumn{1}{c|}{PSNR}           & MD              \\ \hline
Proposed Strategy           & \multicolumn{1}{c|}{\textbf{45.31}} & \textbf{0.1015} & \multicolumn{1}{c|}{\textbf{50.81}} & \textbf{0.1721} \\ \hline
S1: Mixed Feature Vector    & \multicolumn{1}{c|}{38.80}          & 0.1844          & \multicolumn{1}{c|}{48.94}          & 0.1826          \\ \hline
S2: Addition fusion         & \multicolumn{1}{c|}{40.65}          & 0.1115          & \multicolumn{1}{c|}{49.73}          & 0.1753          \\ \hline
S3: Single 3D Grid only     & \multicolumn{1}{c|}{42.86}          & 0.1518          & \multicolumn{1}{c|}{50.32}          & 0.1914          \\ \hline
S3: Three 2D Planes only    & \multicolumn{1}{c|}{32.91}          & 0.2115          & \multicolumn{1}{c|}{45.29}          & 0.2001          \\ \hline
S4: 2D Plane for parameters & \multicolumn{1}{c|}{42.44}          & 0.1365          & \multicolumn{1}{c|}{48.49}          & 0.2002         
\end{tabular}%
}
\end{table}

\section{Ablation Study}
In the main text, four strategies are implemented to enhance model performance and memory efficiency. This section presents an ablation study of these proposed strategies, with results shown in \cref{tab:ablation_study}.
\textbf{Strategy 1:} The proposed approach independently extracts spatial feature vectors and simulation parameter feature vectors before concatenating them. In our ablation study, we apply the Hadamard product to fuse these vectors rather than concatenation, which requires them to share the same dimensionality. We evaluated vector dimensions of 16, 32, and 64, with the optimal results (achieved using 64-dimensional vectors) reported in \cref{tab:ablation_study}.
\textbf{Strategy 2:} While our approach utilizes the Hadamard product to fuse spatial feature vectors with simulation parameter feature vectors, the ablation study examines the effectiveness of simple addition as an alternative fusion method.
\textbf{Strategy 3:} Our proposed spatial encoding method combines a 3D grid with three 2D planes. The ablation study evaluates the performance of using either the 3D grid alone or the 2D planes alone. The 2D planes alone is equivalent to K-Planes.
\textbf{Strategy 4:} Explorable INR encodes each input simulation parameter through a 1D line. In the ablation study, we investigate the use of high-dimension grids, such as 2D planes, for simulation parameters.
The results presented in Table \ref{tab:ablation_study} demonstrate that all proposed strategies contribute to performance improvement.

\section{Loss Function for Parameter Exploration}
\begin{figure}[htbp]
  \centering 
  \includegraphics[width=0.9\columnwidth]{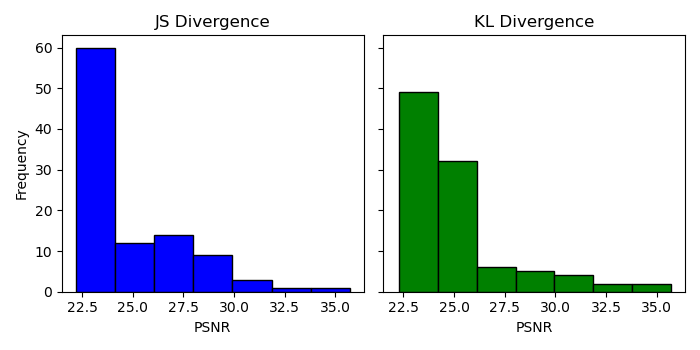}
  \caption{%
    The comparative analysis of KL divergence and JS divergence as loss functions for parameter exploration. We identify 100 outputs and evaluate them using PSNR to assess the effectiveness of each approach.
  } 
  \label{fig:am_hist}
\end{figure}

We employed activation maximization to identify parameters yielding desired distributions. In this section, we evaluate both Kullback–Leibler (KL) divergence and Jensen–Shannon (JS) divergence as potential loss functions. We identify 100 locations with parameter settings through these two loss functions.  Both metrics effectively identified relevant parameters as shown in \cref{fig:am_hist}. Subsequent validation involved predicting volumetric fields using the identified parameters and comparing them to target distributions using PSNR. The comparable performance between KL and JS divergence suggests that both metrics suit our methodology.

\begin{figure*}
  \centering
  \includegraphics[width=1\linewidth]{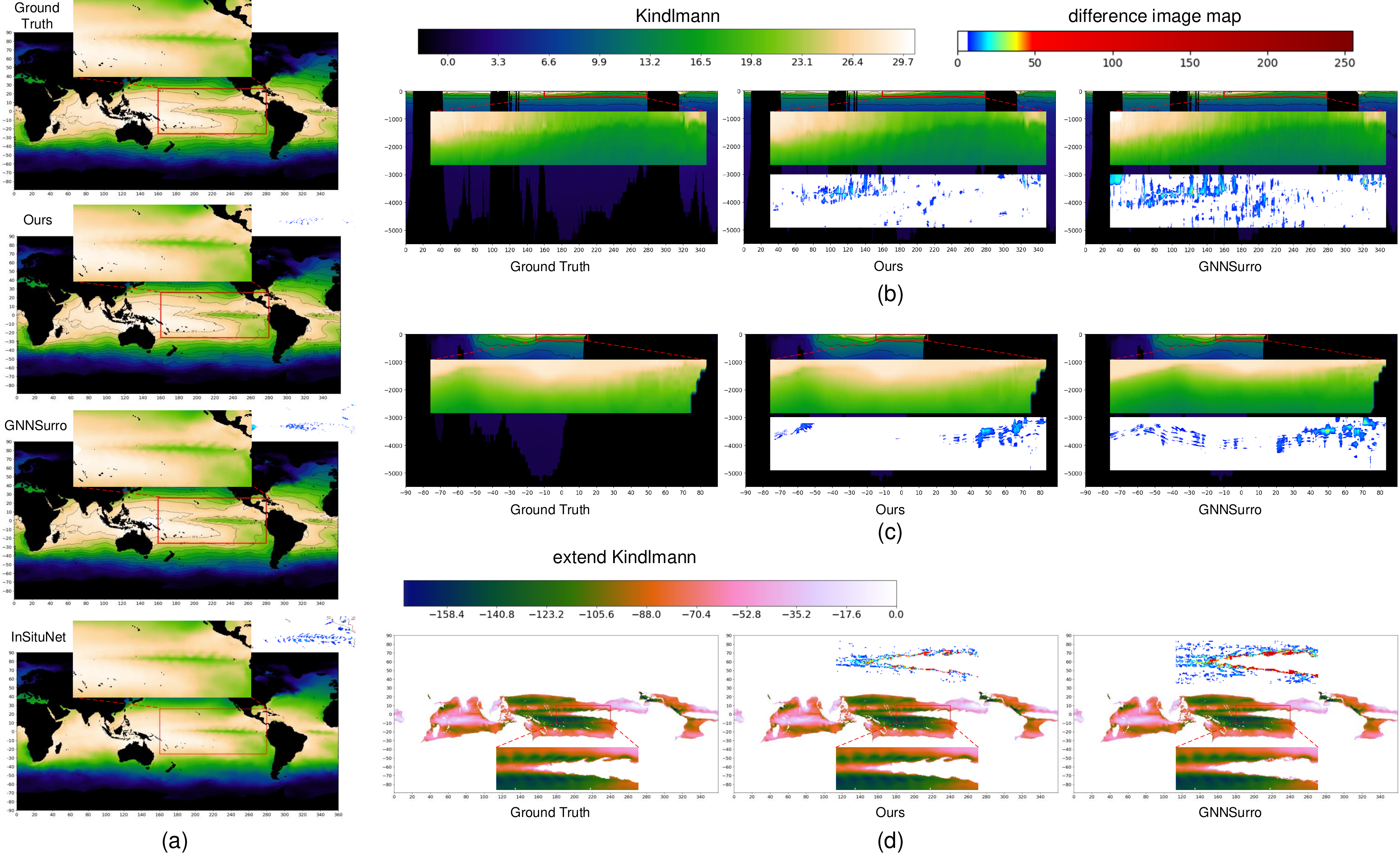}
  \caption{ (a) Comparison of the sea level temperature maps generated using Explorable INR, GNN-Surrogate, and InSituNet, juxtaposed with ground truth maps. 
  Comparison of the vertical cross-sections at (b) the equator (c) $75^\circ E$ generated using Explorable INR and GNN-Surrogate against the ground truth cross-sections. 
  (d) Comparison of the isothermal layer (ITL) depth maps, with a temperature isovalue of $25^\circ C$ generated using Explorable INR and GNN-Surrogate relative to the ground truth. }
  \label{fig:map}
\end{figure*}

\begin{figure*}
  \centering
  \includegraphics[width=\linewidth]{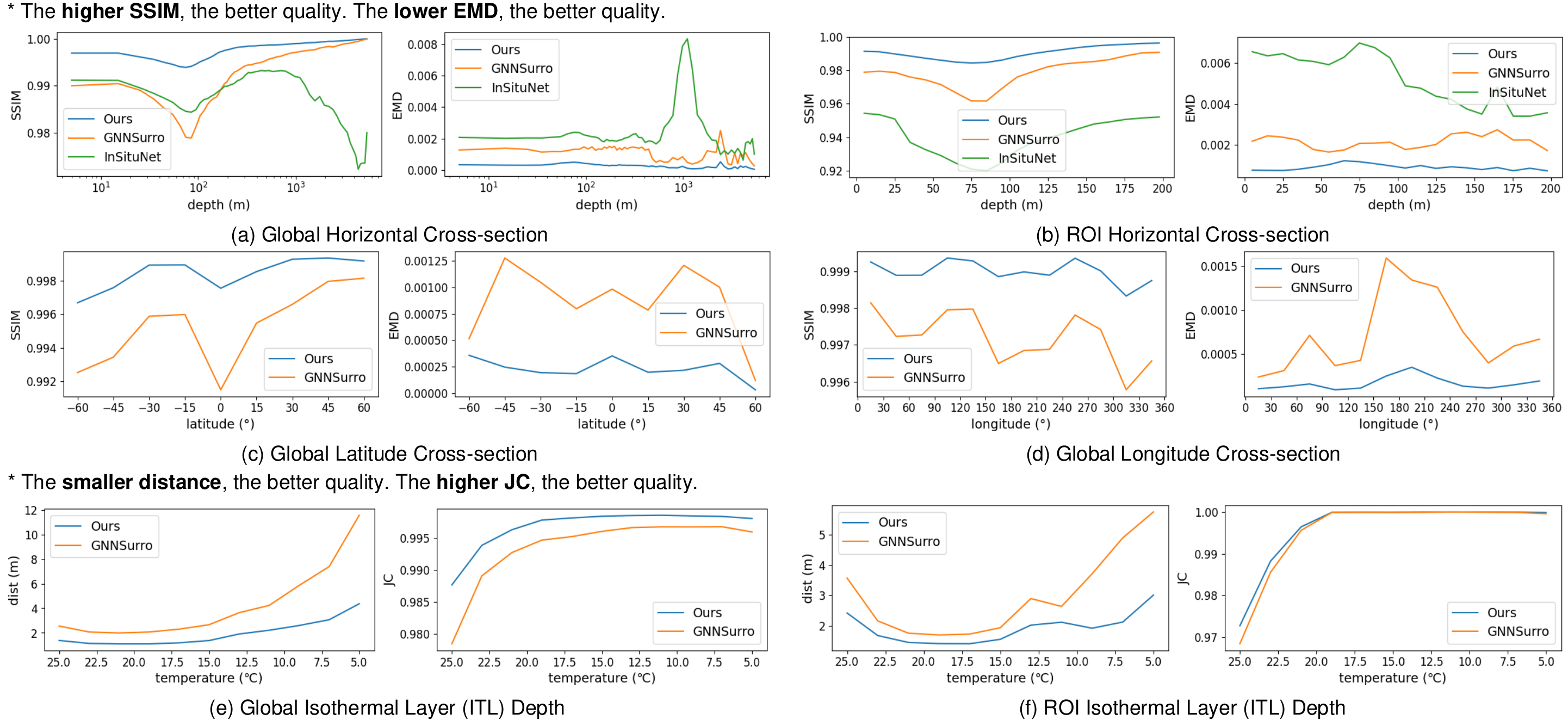}
  \caption{
   A comprehensive evaluation of Explorable INR's performance against GNN-Surrogate and InSituNet through various metrics.
  (a-b) SSIM and EMD across horizontal temperature cross-sections at varying depths.
  (c-d) SSIM and EMD across vertical cross-sections at different latitudes and longitudes. 
  (e-f) Average distance and Jaccard Coefficient (JC) for the depth maps of the isothermal layer (ITL) corresponding to various temperature isovalues. }
  \label{fig:mpas_chart}
\end{figure*}

\section{Detailed Evaluation of the MPAS-Ocean Results}

This section presents a comprehensive evaluation of the MPAS-Ocean results, both quantitatively and qualitatively.

\subsection{Evaluation Metrics}
\textbf{Geometry-level metrics} \quad
The depth of the isothermal layer (ITL) serves as an indicator of the local ocean temperature and its spatial variability, viewed from a geometric perspective.
In this study, we assess the accuracy of computed ITL depths by evaluating their overlap using the Jaccard coefficient and measuring the mean surface distance within intersecting regions.

\textbf{Image-level metrics} \quad
For the analysis at the image level, both horizontal and vertical cross-section images were rendered, where the depth, latitude, or longitude were fixed.
We employed the Kindlmann colormap~\cite{kindlmann2002face} to map colors to ocean temperatures, establishing a visual correspondence.
To quantify the similarity between two rendered images, we utilized the Structural Similarity Index Measure (SSIM) and the Earth Mover’s Distance (EMD) between color histograms~\cite{he2019insitunet}.

Our evaluation encompassed both global perspectives and specific regions of interest (ROI).
For the ROI analysis, we selected a region ranging from $160^\circ W$ to $80^\circ E$, $26^\circ S$ to $26^\circ N$, the sea surface down to a depth of 200 meters. 
This area corresponds to the eastern equatorial Pacific cold tongue.
Regarding image-level evaluation, the resolution varied according to the targeted region: $1024 \times 512$ for global images and $420 \times 180$ for ROI-focused images.

\subsection{Quantitative and Qualitative Analysis Results}

The evaluation results are organized into two main categories:
(1) geometry-level analysis, as illustrated in \cref{fig:map}(d) and \cref{fig:mpas_chart}(e, f); and 
(2) image-level analysis, depicted through  \cref{fig:map}(a-c) and \cref{fig:mpas_chart}(a-d).

\textbf{Geometry-Level analysis} \quad
Temperature isovalues ranging from $25^{\circ}C$ to $5^{\circ}C$ were sampled to calculate the depth of the isothermal layer (ITL). 
The quantitative results are illustrated in \cref{fig:mpas_chart} (e, f).
In this figure, temperatures are organized in descending order to reflect the ocean's typical monotonic decrease in temperature with increasing depth, whereby a higher temperature isovalue corresponds to a shallower depth.
Explorable INR demonstrates superior performance with a smaller mean surface distance and greater surface overlap (as measured by the Jaccard Coefficient) compared to the GNN-Surrogate.

\cref{fig:map}(d) showcases a comparison of the $25^{\circ}C$ isothermal layer (ITL) depth maps generated by Explorable INR and GNN-Surrogate, utilizing the Extended Kindlmann colormap~\cite{moreland2016we} for visualization.
Explorable INR produces a depth map that more closely aligns with the ground truth, particularly in capturing the jagged details in the gap across the equator in the eastern Pacific.
This ITL depth map comparison raises an intriguing question about the origins of the observed gap, which is subsequently addressed through image-level analysis.

\textbf{Image-Level analysis} \quad
In addition to utilizing the Kindlmann colormap for rendering, inspired from prior research~\cite{shi2022GNN}, this study incorporates difference images to highlight significant pixel variations (where the difference $\triangle \geq 6.0$ in the CIELUV color space), drawing inspiration from prior research.

We rendered different horizontal cross-sections from the sea surface to the seabed. 
In \cref{fig:mpas_chart}(a), Explorable INR outperforms GNN-Surrogate and InSituNet, evidenced bu higher SSIM and lower EMD, both globally and within the ROI. 
\cref{fig:map}(a) shows the sea level rendering results. 
The zoom-in views and the different images reveal that Explorable INR more accurately reflects the ocean temperature compared to  GNN-Surrogate and InSituNet.
Note that as InSituNet directly predicts the images, it does not facilitate the straightforward incorporation of isotherms for enhanced visualization.

This study also includes vertical cross-sections of the temperature field for analysis.
In \cref{fig:mpas_chart}(c, d), Explorable INR is shown to produce images with higher SSIM and lower EMD compared to those generated by GNN-Surrogate.
Qualitatively, in \cref{fig:map}(b), the vertical cross-section from the equator, and \cref{fig:map}(c), a vertical cross-section from $75^\circ E$, it is shown that Explorable INR's predictions yield a smaller difference from the ground truth than those of the GNN-Surrogate.

\begin{figure}[p]
  \centering 
  \includegraphics[width=\columnwidth]{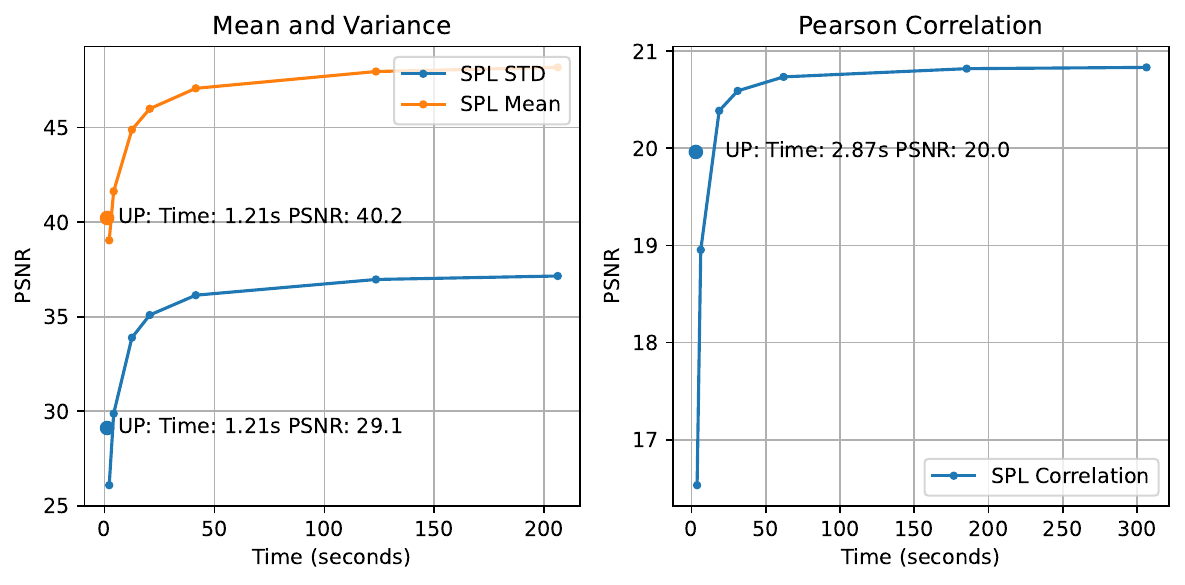}
  \caption{%
    Left: the figure for mean and variance field. Right: the figure for the correlation field. The line plot shows the relationship between the sampling time and PSNR for the SPL method and the dot shows the result for the UP approach. 
  } 
  \label{fig:mean_var_mpas}
\end{figure}

\section{Spatial Exploration for MPAS-Ocean}

We conducted a similar experiment on the MPAS-Ocean dataset for the mean, variance, and covariance fields across ensemble members. The ground truth was derived from the original 100 ensemble runs. \cref{fig:mean_var_mpas} illustrates the PSNR of uncertainty propagation (UP), and the sampling method (SPL) compared to the ground truth fields.
The left figure shows the results for the mean and variance fields, while the right panel is for the correlation field. UP achieves faster computation times than the sampling method, even when using a small number of samples (five). Moreover, UP's quality slightly surpasses that of the sampling method with five samples. The trends in accuracy and computation time align closely with our observations from the Nyx dataset experiments.

\begin{figure}[htbp]
  \centering 
  \includegraphics[width=.78\columnwidth]{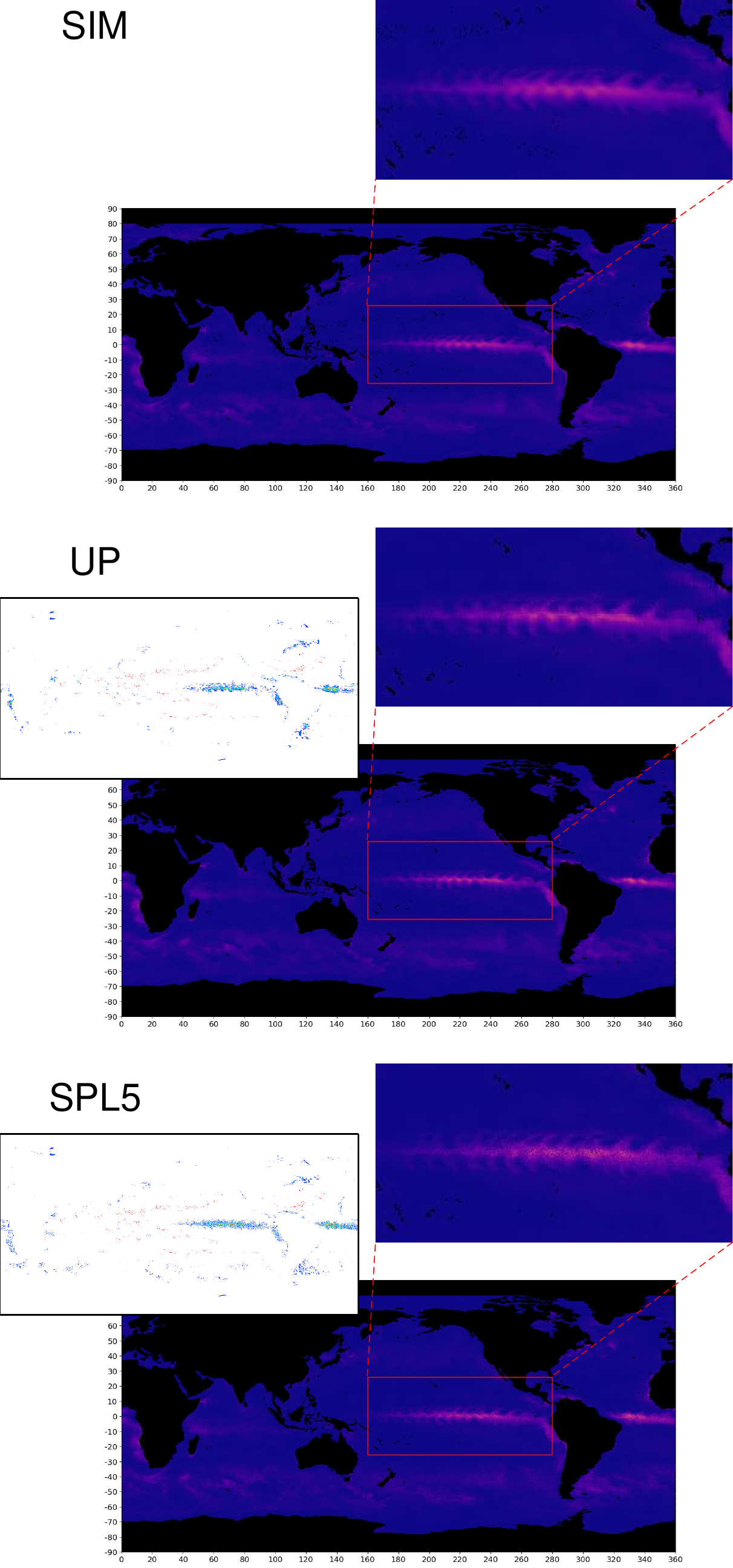}
  \caption{%
  The qualitative comparison of the variance field. SIM stands for the variance filed computed from the 100 simulation runs. UP stands for uncertainty propagation. SPL5 stands for the sampling method with five samples. For UP and SPL5, we also show the error maps on the left. 
  } 
  \label{fig:mpas_variance}
\end{figure}

\cref{fig:mpas_variance} presents a qualitative analysis. The UP approach generally yields smaller errors than the sampling method. Furthermore, UP shows superior computational efficiency, requiring only 1.21 seconds compared to 2.17 seconds for 5 samples.
The results show the effectiveness of our UP approach in balancing accuracy and computational efficiency when analyzing ensemble data from complex ocean simulations.

\section{User Opinion}
The correlation analysis is based on the domain scientists' interest in the statistics of the ensemble dataset and the relation between any locations. To validate the alignment of this analytical approach with the perspective of scientists, we interviewed a domain scientist.
Through the consultation, we identified several valuable applications of correlation analysis. One primary application is cost optimization, particularly in scenarios where comprehensive real-world data collection is expensive or logistically challenging. The correlation field enables scientists to strategically select reference points that have lower costs and strong correlations with target points, thereby optimizing data collection efforts. Additionally, correlation analysis serves as an indicator of the sensitivity of a surrogate model. Our method helps efficient queries within designated ranges of a given parameter, allowing scientists to evaluate the performance of the model across individual parameters and establish confidence in its predictive capabilities.

\bibliographystyle{abbrv-doi-hyperref}

\bibliography{supplementary}

%% file: tex/1_introduction.tex
With the advancement of supercomputing technologies, high-resolution ensemble simulations in domains such as cosmology and oceanology allow scientists to investigate complex physics-based systems.
However, running ensemble simulations requires not only vast amounts of computing hours to generate results but also additional resources to load, analyze, and visualize them.
To overcome computational and memory constraints, recent research has introduced several surrogate models designed to facilitate ensemble simulation analysis and exploration of the parameter space. Examples include data-based models like NNVA \cite{hazarika2019nnva}, GNN-Surrogate \cite{shi2022GNN}, and VDL-Surrogate \cite{shi2022vdl}, as well as image-based approaches such as InSituNet \cite{he2019insitunet}.
Despite advancements, domain scientists still face two challenges in effectively using these models to explore ensemble simulations.

\textbf{Challenges} \textbullet \, \textbf{C1: Spatial Exploration} Domain scientists often aim to gain an overview of the ensemble by examining the physical attribute distribution across the members at particular spatial locations and identifying spatial relationships within the dataset. 
However, image-based approaches are limited as they predict rendered images rather than the raw data, hindering field analysis across different parameter settings.
Meanwhile, existing data-based models are designed to generate the entire field for a given parameter setting.
As a result, obtaining overviews in local regions requires consecutive steps including 1) sampling from the simulation parameter sub-domain, 2) predicting the entire field for each sample, and 3) subsequently conducting statistical summaries. These steps are both memory- and computation-intensive. 
\textbullet \, \textbf{C2: Parameter Exploration} In many scientific applications, instead of aiming to predict the physical features for given simulation parameters, the goal is to identify simulation parameters that produce desired distributions, which is known as an inverse problem. 
While traditional approaches to these problems often involve training neural networks as surrogate models for faster data generation compared to numerical simulations, the vast search space in these applications can make the rendering computationally expensive even with advanced sampling methods.

To address these challenges, we propose an explorable ensemble surrogate model based on implicit neural representations (INR), named \textit{Explorable INR}, which enables point-based spatial query for a given set of parameters without the need to compute the full-scale field data. 
INRs are a class of neural networks that learn continuous functions to represent data, bypassing the need for explicit grids or structured data inputs, such as images \cite{sitzmann2020implicit,luo2024hierarchical} or scalar fields \cite{lu2021compressive, weiss2022fast,luo2024continuous}.
Specifically, for volumetric data, INR maps 3D coordinates to the corresponding physical features such as density or temperature. 
Our INR-based surrogate model enables scientists to query data at a given location and parameter, facilitating the exploration of regions of interest in both spatial domains and parameter domains with reduced memory and computational costs compared to existing surrogate models.
However, enabling the point query is not enough to solve the challenges since both challenges involve either obtaining or matching the distribution that requires querying the INR model across multiple points within the domain. This operation becomes increasingly burdensome as the dimensionality of the problem grows. 
Thus, we also propose the following \textit{model-agnostic} approaches to facilitate the exploration of ensemble simulations through INR.

\textbf{Solutions.} Specifically, to address \textbf{C1}, we adapted a functional-based method that approximates the statistical summary of a region of interest efficiently.
Li and Shen \cite{li2024improving} have already demonstrated that efficient acquisition of statistical value distributions across an input region can be applied to INRs through uncertainty propagation. 
Specifically, the nonlinear transformations applied by INR can shift and distort the underlying data distribution, posing challenges in preserving the statistical integrity of the region. 
To resolve this, a Probabilistic Affine Form (PAF) is employed to propagate the parameter and spatial distribution from the input to the value distribution in the INR output.
In this paper, we demonstrate that by integrating uncertainty propagation with the feature grid-based INRs, previously infeasible or costly ensemble analysis and visualization can be achieved efficiently with Explorable INR.
To address \textbf{C2}, we reformulate the inverse problem as an optimization task. Specifically, we minimize the Kullback-Leibler (KL) divergence between the desired distribution and the distribution of INR predictions.
The desired distribution is from a given set of simulation parameters, while the distribution of INR predictions is from a random set of simulation parameters.
Inspired by NNVA \cite{hazarika2019nnva}, we subsequently apply gradient descent to find one or more simulation parameters that align with the given distribution.
In this work, we show that the gradient descent method utilized in NNVA is also effective for uncertainty propagation and PAF, which enables a more effective and scalable approach to solving inverse problems in scientific applications.

In summary, the contributions of our work are as follows:
\begin{itemize}
\item We tailor the implicit neural representation (INR) to develop an explorable ensemble surrogate model that exhibits significantly lower computation and memory cost compared to existing neural-network-based models.

\item We improve the probabilistic affine form (PAF) in the uncertainty propagation to represent the uncertainty in the feature-grid domain. We also illustrate that linear dependencies between PAFs can be efficiently calculated to allow auto-correlation analysis with the Explorable INR.

\item We showcase the application of gradient descent to uncertainty propagation on the Explorable INR, employing this method for effective parameter space exploration.
\end{itemize}

%% file: tex/2_related_work.tex
\section{Related Works}


\subsection{Implicit Neural Representation}

Implicit Neural Representations (INRs), which employ coordinate-based multi-layer perceptrons (MLP) to represent field data, have recently emerged as a hot research topic. 
Xie et al. \cite{xie2022neural} provide an extensive summary of related applications and architectures of INRs. 
In this section, we review related works that address overcoming spectral bias and applying INRs in scientific visualization.

INRs, i.e. coordinate-based networks, face difficulties in learning functions due to low-frequency spatial input \cite{rahaman2019spectral}, a phenomenon known as spectral bias. 
To address this issue, researchers have proposed various methods, including positional encoding through high-frequency signals \cite{zhong2019reconstructing} and periodic activation functions \cite{sitzmann2020implicit}.
Recently, parametric encoding has emerged as a promising approach, offering high efficiency and superior performance.
This method combines trainable parameters with auxiliary data structures to interpolate parameters based on input coordinates.
The auxiliary data structure can be grids \cite{muller2022instant, fridovich2022plenoxels, fridovich2023k, chen2022tensorf, wurster2024apmgsrn, xiong2024regularsrn} or trees \cite{takikawa2021neural}.
In this paper, we propose a novel grid-based INR for parametric encoding of ensemble surrogate models.

In scientific visualization, the first category of INR applications is reducing storage costs. For 3D spatial data, Lu et al. \cite{lu2021compressive} demonstrate that INR-based methods can achieve better compression rates than traditional algorithms, but with longer latency. Some later studies \cite{weiss2022fast, wu2024interactiveVolVis} addressed this latency issue by using  GPU on-chip memory.
The second category involves using INRs to reconstruct 3D data from their 2D projections. Sitzmann et al. \cite{sitzmann2019scene} proposed a scene representation network to learn the 3D continuous field from 2D observations. NeRF (Neural Radiance Fields) \cite{mildenhall2021nerf} is one of the pioneering works in this area, representing a scene as a neural radiance field. Given a set of images with known camera poses, NeRF can synthesize images at novel viewing angles using the learned radiance field, demonstrating the potential of INRs in reconstructing 3D information from 2D data.
Similarly, in medical imaging field, INRs are used to estimate density fields from restricted viewing angles and sparse observations, converting predicted densities to sensor domains via Fourier transforms for MRI scans or Radon transforms for CT scans \cite{zang2021intratomo, shen2022nerp}. These applications demonstrate the versatility of INRs in reconstructing detailed internal structures from limited external data.
The third category is using INR to reduce the computation. Some Physics-informed neural networks \cite{raissi2019physics} adapt INR models that are augmented with additional loss functions enforcing physical rules, or introduced to reduce the computational demands of simulations. Many other studies have extended this idea to various types of partial differential equations \cite{zehnder2021ntopo, izzo2022geodesy, smith2020eikonet, pfrommer2021contactnets} that describe different physical phenomena. These approaches demonstrate the potential of INRs to incorporate physical constraints into the learning process.
The usage of INRs in this paper falls into both the first and the third categories, where we use the INR as a surrogate model to reduce both the storage and computation costs.

\subsection{Ensemble Simulation Analysis and Visualization Tasks}



\textbf{Ensemble Overview.} Ensemble overview is a prevalent task essential for almost all analyses of ensemble simulations. The goal is to present all members and reveal the collective behavior\cite{wang2018visualization}, which can be spatial or temporal. Statistical distributions for ensemble members are required to present a comprehensive overview. Various techniques can be employed to visualize the statistical summary through the distribution, including box plots \cite{whitaker2013contour, mirzargar2014curve}, probabilistic iso-surfaces \cite{pothkow2010positional, pothkow2011probabilistic}, and quantile trend charts \cite{potter2009ensemble}.
Our spatial domain exploration with Explorable INR efficiently obtains distributions across ensemble members at any spatial position. These distributions can then be visualized using the above techniques to provide a comprehensive ensemble overview.

\textbf{Spatial and Temporal Analysis.} The spatial and temporal analysis of the ensemble data involves comparing the spatial difference or the temporal trends among a collection of values (ensemble member values). This analysis can be achieved by first obtaining the statistical distribution of ensemble values at a specific spatial or temporal position. Subsequently, techniques such as clustering \cite{kniss2003gaussian, liu2016comparative, wang2016multi}, correlation calculation \cite{farokhmanesh2023neural, pfaffelmoser2012visualization, evers2021uncertainty}, or trend plots \cite{hollt2014ovis, wang2018visualization} can be applied to explore the spatial relationships and temporal trends within the ensemble data.

This paper focuses on metrics indicating spatial or temporal relations in ensemble data. For clustering grid points, we calculate distance metrics between distributions rather than single values. A previous work \cite{love2005visualizing} used KL-divergence for this purpose.
Pearson correlation and mutual information are commonly used to understand spatial and temporal dependencies in ensemble data, particularly in meteorology. A previous study \cite{farokhmanesh2023neural} employed a specialized neural network to predict these dependencies, addressing computational and memory challenges.
Our Explorable INR model with uncertainty propagation avoids explicit correlation prediction or dense data reconstruction. By representing spatial value distributions as probabilistic affine forms, we efficiently calculate Pearson correlation coefficients between positions, enabling deeper insights into spatial or temporal relationships in ensemble data.



\textbf{Parameter Analysis.} In ensemble simulations, a single parameter setting is associated with a spatial-temporal field. Various previous works have attempted to visualize this connection. In these studies \cite{biswas2016visualization, kehrer2010interactive}, multiple views are often combined to simultaneously display the parameter space and the field data. In recent research, neural-network-based surrogate models are employed to analyze the sensitivity of parameters on the simulation results \cite{he2019insitunet, shi2022vdl} and to suggest desirable parameters for a target simulation output \cite{hazarika2019nnva}. These tasks are crucial for domain scientists, as they enable the use of surrogate models, which require significantly fewer resources than running full simulations, to identify potential parameter settings for further simulation and analysis. By leveraging these surrogate models, scientists can efficiently explore the parameter space and gain insights into the relationships between parameters and simulation outcomes. Our Explorable INR is also capable of parameter sensitivity analysis and parameter suggestion. In addition, the ``target simulation output'' for parameter suggestion is not limited to a specific scalar field or minimizing/maximizing the output value. The target can also be a scalar value distribution, which provides more flexibility in describing the desired features of interest.

%% file: tex/3_method.tex
\section{Methodology}
\label{sect:method_inr}

\begin{figure*}
\centering
\includegraphics[width=\textwidth]{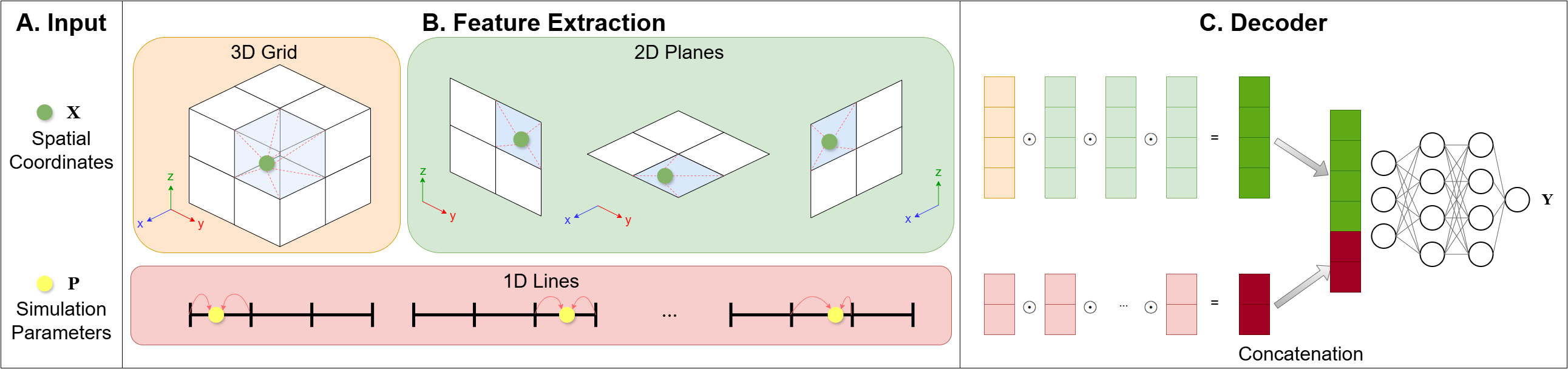}
\caption{
The proposed Explorable INR works as follows: (A) Take coordinates and simulation parameters as input. (B) Query spatial coordinates in XYZ 3D feature grid ($64^{3}$), and XY, YZ, and XZ feature planes ($256^{2}$); query simulation parameters on 1D feature lines ($16^{1}$). The features are interpolated by corner features. (C) The spatial and parameter features are fused via the Hadamard product; the fused vectors are concatenated into an ensemble feature, which is then decoded by an MLP with 3 hidden layers and 128 hidden nodes to predict the physical feature.} 
\label{fig:pipeline}
\end{figure*}

\textbf{Problem Statement.} An ensemble simulation comprises a collection of simulation runs, each with the same initial condition but varying parameter settings $\textbf{P}$, facilitating the exploration of potential outcomes and inherent stochasticity within the model.
$\textbf{P} \in \mathbb{R}^{m}$ are the $m$ simulation parameters.
Such ensemble simulation can be represented as a functional mapping from simulation parameters to the output volumetric data, where the output consists of $N$ data points, each defined by a coordinate and value pair $(\textbf{X}, \textbf{Y})$.
Here, $\textbf{X} \in \mathbb{R}^{d}$ represents the coordinates in a $d$-dimensional space, and $\textbf{Y} \in \mathbb{R}^{n}$ represents the value in $n$-dimensional space, with $d$ typically being either 3 or 4 to represent non-time-varying or time-varying data, and $n$ being 1 or 3 for scalar or vector fields, respectively.
In this work, we focus on a time-independent three-dimensional scalar field within the ensemble simulation framework.
Generally, these value pairs are produced at grid vertices, and values at any arbitrary point can be interpolated.
However, storing high-resolution data for accurate interpolation requires substantial disk space.
To mitigate storage requirements, researchers have explored implicit neural representations to learn the functional mapping from coordinate $\textbf{X}$ to value $\textbf{Y}$.
Despite advances, employing a single INR for each simulation output does not meet the requirement of scientists who need to explore various simulation parameter settings, given the extensive time required to train INRs across multiple simulation parameter settings and the resulting substantial storage burden.
To address these challenges, we developed an INR to learn the function \cref{eq_model} that maps both the spatial coordinates and the simulation parameters to the output volume.
\begin{equation} \label{eq_model}
    F(\textbf{X}, \textbf{P}) = \textbf{Y},
\end{equation}
Consequently, our model is named \textit{Explorable INR}.

\textbf{Model Architecture.}  In this work, we employ a hybrid approach that combines feature grids and planes with a smaller MLP decoder, inspired by Instant-NGP~\cite{muller2022instant} and K-planes~\cite{fridovich2023k}.
This approach, in contrast to purely MLP-based methods that utilize a multilayer perceptron with a periodic activation function, as discussed in SIREN~\cite{sitzmann2020implicit} and Neurcomp~\cite{lu2021compressive}, significantly reduces the training time.
The increased efficiency makes a hybrid method practical that adopts complicated mapping from both spatial and parameter domains.

The architecture of our Explorable INR model, illustrated in \cref{fig:pipeline}, processes spatial coordinates and simulation parameters through a multi-stage pipeline. 
Initially, both inputs go through feature encoding (\cref{fig:pipeline} (B)) using learnable feature grids. 
In this process, the domain is discretized into a grid where each vertex possesses a learnable feature vector. 
For any query point, its feature vector is derived through linear interpolation of the surrounding cell vertices. 
Spatial encoding utilizes XYZ coordinates for a 3D grid and XY, YZ, and XZ for 2D planes, represented as $f^{g}{sp}(\textbf{X})$, where $g \in G$, and $G$ encompasses the 3D grid and 2D planes. 
Parameter encoding employs individual parameters to query 1D lines, denoted as $f^{l}{p}(\textbf{P})$, where $l \in L$, and $L$ is the set of 1D lines.
Subsequently, feature fusion (\cref{fig:pipeline} (C)) occurs through Hadamard product operations:
\begin{equation} \label{eq_sp}
    F_{sp}(\textbf{X}) = \odot_{g \in G} f^{g}_{sp}(\textbf{X}),
\end{equation}
\begin{equation} \label{eq_p}
    F_{p}(\textbf{P}) = \odot_{l \in L} f^{l}_{p}(\textbf{P}),
\end{equation}
where $\odot$ denotes the Hadamard product applied to the sequence of features.
The fused spatial and parameter feature vectors are then concatenated to form an ensemble feature vector. 
Finally, a shallow MLP decodes this ensemble feature vector and yields the physical feature output.

In the design of the Explorable INR, we utilized several key strategies to enhance model performance and memory efficiency. 
\textbf{Strategy 1} The feature vectors for spatial coordinates and simulation parameters are independently extracted.
This approach is rooted in the hierarchical nature of ensemble simulations, where simulation parameters determine the entire field before spatial coordinates are queried. 
Moreover, this design aligns with the concept of conditional INRs, where the feature vector of the simulation parameter serves as a learnable condition vector.
In addition, this strategy avoids memory complexity issues of standard multi-resolution grids \cite{muller2022instant} that are crucial for large scientific datasets because the standard multi-resolution grid will be extended to high dimensional grids with memory complexity $\mathcal{O}(2^{m+d})$.
\textbf{Strategy 2} We adopted the Hadamard product as the fusion operator for both spatial and simulation parameter feature vectors. 
This choice was informed by research conducted by Fridovich-Keil et al. \cite{fridovich2023k}, which demonstrated the superiority of element-wise multiplication over addition in feature vector fusion.
\textbf{Strategy 3} We developed a hybrid spatial feature extraction approach, combining low-resolution 3D feature grids with high-resolution 2D feature planes.
This method balances the performance and memory efficiency, addressing memory constraints by the low-resolution 3D grid while enhancing spatial details by the high-resolution 2D planes.
3D Grid from Instant-NGP \cite{muller2022instant} is a great encoding scheme to represent low-frequency signals well, while K-Planes \cite{fridovich2023k} utilizes few parameters to give high-frequency detail at fine resolutions. Combining them can lead to the highest parameter efficiency for accuracy with high throughput.
\textbf{Strategy 4} One-dimensional feature lines are employed for the feature extraction of simulation parameters.
This strategy effectively mitigates the risk of memory explosion that could arise with a significant increase in parameter dimensions. 
The reason for this approach is twofold: first, it assumes parameter independence, contrasting with the potentially interconnected nature of spatial domain data; second, data points in the parameter domain tend to be sparser than those in the spatial domain, making a high-dimensional grid less likely to achieve convergence during training.
These strategies enable the Explorable INR to efficiently handle complex and large-scale ensemble simulation datasets, striking a balance between model accuracy and computational resource management. To prove the effectiveness of these strategies, we conduct an ablation study in the appendix.

\textbf{Remark 1} \textit{The proposed Explorable INR outperforms the K-planes model due to its higher-degree polynomial feature vector, allowing it to capture more intricate details from the training data. A detailed justification can be found in the appendix.}

\subsection{Spatial Exploration}
\label{sect:spat_explore}
\textbf{Setup.} As mentioned in the introduction, a major visual analysis task for ensemble simulations is to obtain the physical feature distribution of a specific spatial location across ensemble members.
For instance, climate scientists may be interested in the average temperature of a location this month, rather than the temperature in one day.
However, enabling point query for the Explorable INR is not efficient enough because of the time-consuming nature of the dense sampling involved in the computation of distributions.
Hence, a shift from \textit{point estimation} to \textit{regional estimation} is anticipated:
Specifically, from \cref{eq_model}, INRs focus on estimating $F(\textbf{X}_{i}, \textbf{P}_{j})$ at specific points $\textbf{X}_{i} \in \mathbb{R}^{d}$, $\textbf{P}_{j} \in \mathbb{R}^{m}$.
Assume the region of interest in simulation parameter space is $\mathcal{R} \subset \mathbb{R}^{m}$.
The regional estimation involves calculating
\begin{equation} \label{eq_region_integral1}
    \hat{F}(\textbf{X}, \mathcal{R}) = \frac{1}{|\mathcal{R}|} \int_{\mathcal{R}} F(\textbf{X}, \textbf{P})d\textbf{P}.
\end{equation}

To solve \cref{eq_region_integral1}, traditional methods based on Monte Carlo or grid-based sampling require a large number of samples $\mathbf{P}_j$ to accurately approximate the solution over the domain $\mathcal{R}$. In a many-query context, this leads to a computationally expensive process, with the overall cost scaling as $\mathcal{O}(N_{\textbf{P}})$, where $N_{\textbf{P}}$ is the number of samples for simulation parameters. As the dimensionality of simulation parameters increases, the number of samples required to maintain a given sampling density grows exponentially, a phenomenon known as the ``curse of dimensionality'' in sampling theory.

In this paper, we propose an efficient approximation method that leverages regional estimation to reduce sample requirements, enabling accurate predictions with fewer computations.

\textbf{Mean Estimation.}
In ensemble simulations, the scalar or vector values at a specific spatial location across different ensemble members can be treated as a random variable. 
This is due to the inherent uncertainties in the simulation process, which lead to variations in output values across ensemble members. 
However, deriving statistical distributions of this random variable from INR models requires inference with all possible parameter settings.
While the inference cost is not high in modern feature grid-based INR, the computational cost escalates rapidly as the complexity of the model increases, such as additional feature grids or larger feature decoders.
To address this potential challenge, we have adopted the uncertainty propagation method proposed by Li and Shen \cite{li2024improving}. 
The core concept involves representing parameter ranges as an affine combination of random variables, referred to as a probabilistic affine form (PAF), which can be propagated through the neural network like the forward pass of an input vector. 
The output PAF characterizes the distribution of the physical attribute across the ensemble simulations.
In the subsequent sections, we will elaborate on how to perform uncertainty propagation to our Explorable INR model.

Our Explorable INR takes two vectors as input: $\textbf{X} \in \mathbb{R}^{d}$ and $\textbf{P} \in \mathbb{R}^{m}$.
As described in \cref{sect:method_inr}, $\textbf{X}$ is a spatial coordinate, while $\textbf{P}$ denotes a parameter setting, where $m$ is the number of simulation parameters.
In ensemble simulation, the parameters of interest are characterized by a random vector $\hat{\textbf{P}} \in \mathbb{R}^{m}$, whose distribution is typically provided by domain scientists.
If no prior knowledge is given regarding the parameter domain, we assume that parameters are independently and uniformly sampled within a range between $p_{-}$ and $p_{+}$ for each parameter, where $p_{-}$ and $p_{+}$ are given by the user.
The value distribution at a specific position $\textbf{x}_0$ across the possible parameter settings $\hat{\textbf{P}}$ is represented by the output random vector $\textbf{Y}_{\textbf{x}_0, \hat{\textbf{P}}}$.
To apply uncertainty propagation, the first step is to represent the input random vector $\hat{\textbf{P}}$ in the PAF.

\begin{figure}[ht]
 \centering 
 \includegraphics[width=0.9\columnwidth]{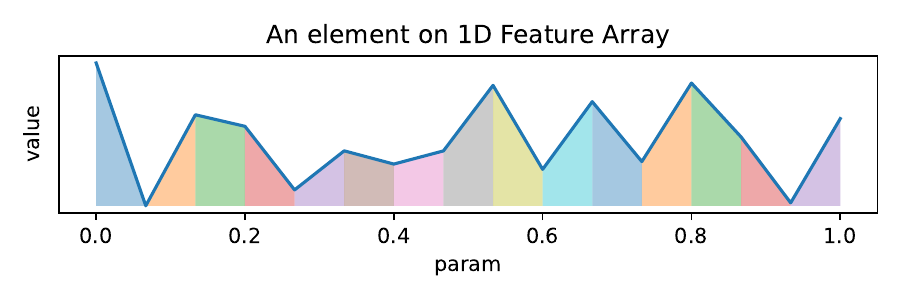}
 \caption{The plot of interpolated feature element value against input parameter is a piecewise linear function for a specific parameter. Different color indicates different pieces of the function.}
 \label{fig:piecewise_linear}
\end{figure}

From Li and Shen \cite{li2024improving}, the definition of input ranges in the parameter space is
\begin{equation} \label{P0}
\hat{\textbf{P}} = \textbf{p}_{0} + \sum_{i=1}^{N} \textbf{p}_{i}Z_{i},
\end{equation} 
where $\textbf{p}_{0}$ denotes the mean vector in the parameter ranges, $\textbf{p}_{i}$ represents an axis-aligned vector in the parameter space, and $Z_{i}$ is a random variable associated with parameter $i$ within the specified parameter range. $Z_{i}$ is a standard unit random variable with $mean(Z_{i}) = 0$ and $var(Z_{i}) = 1$. 
Therefore the value in the axis-aligned vector $\textbf{p}_{i}$ defined the standard deviation for the parameter $i$.
Generally speaking, the mean and variance of an input range can be easily computed, as the range of a single input element typically follows a uniform distribution.
However, in the proposed feature grid-based INR, the distribution is not uniform but rather a superposition of multiple uniform distributions.
This complexity arises from the use of linear interpolation on feature grids.
\cref{fig:piecewise_linear} illustrates how an individual element value from the feature vector varies with respect to the input parameter, resulting in a piecewise linear function $F_{piece}$. 
The domain of this function can be partitioned into disjoint subdomains. 
Consider a subdomain defined on an interval $[a_{1}, a_{2}]$, where $F_{piece}(a_{1}) = v_{1}$ and $F_{piece}(a_{2}) = v_{2}$. 
For this subdomain, we can define a parameterized function $f_{piece1}(x) = v_{1} + (v_{2} - v_{1})x$, where $x \in [0, 1]$.
In more general cases, the endpoints of a range query may not align with the subdomain endpoints and may span multiple disjoint subdomains.
To compute the mean and variance for a given range query, it is crucial to identify both the range query endpoints and all interval endpoints within the query range. 
Given the query range $[a_{1}, a_{n}]$, $a_{2}$ to $a_{n-1}$ are sorted endpoints of subdomains in the query range.
We can then define parameterized functions for each pair of adjacent points, and derive the mean for the given query range.
\begin{equation}
\label{eq:mean}
    Mean = \frac{1}{\sum_{m=1}^{n-1} (a_{m+1}-a_{m})} \sum_{m=1}^{n-1} (a_{m+1}-a_{m}) \frac{v_{m}+v_{m+1}}{2}.
\end{equation}

In the equation, $a_{m+1}-a_{m}$ is consistent for most subdomains except for the range [$a_{1}$, $a_{2}$] and [$a_{n-1}$, $a_{n}$].
The variance can also be calculated:
\begin{equation}
\label{eq:var}
    Var = \frac{1}{a_{n}-a_{0}} \sum_{m=1}^{n-1} (\frac{\mathbf{m}^{2}(a_{m+1}^{3}-a_{m}^{3})}{3} + \mathbf{m}\mathbf{b}(a_{m+1}^{2}-a_{m}^{2}) + \mathbf{b}^{2}(a_{m+1}-a_{m}))
\end{equation}

$\mathbf{m}=(v_{m+1}-v_{m})/(a_{m+1}-a_{m})$ and $\mathbf{b}=v_{m}-\mathbf{m}a_{m}$ are the slope and intercept for each parametrized function.
With the mean and variance of the feature, the uncertainty propagation technique can be applied to the Explorable INR.
The input PAF in the feature space is defined as $\hat{\textbf{P}}_0 = \textbf{p}_{0,0} + \sum_{i=1}^{N} \textbf{p}_{0,i}Z_{i}$, where the values in $\textbf{p}_{0,0}$ are the mean values calculated using \cref{eq:mean} for each feature dimension and each axis-aligned vector $\textbf{p}_{0,i}$ corresponds to variance calculated using \cref{eq:var}.

The PAFs can be propagated through the INR model following the affine arithmetic rules. For linear layers in the network, the affine form is propagated as
\begin{equation}
\label{eq:parameter_linear}
    \hat{\mathbf{P}}_{\text{out, linear}} = \textbf{W}\hat{\mathbf{P}} + \textbf{B} = \textbf{W}\mathbf{p}_{0} + \textbf{B} + \sum_{i=1}^{N} \textbf{W}\textbf{p}_{i}Z_{i},
\end{equation}
where $\textbf{W}$ and $\textbf{B}$ are the weight matrix and the bias vector for this linear layer. 
To handle nonlinear functions in the network, such as activation functions, we employ a two-step process.
First, we apply a linear approximation to these functions, and then we introduce additional random vectors $\mathbf{\gamma}_{i}Z_{i}$ to account for the approximation error.
\begin{equation}
    \label{eq:parameter_nonlinear}
    \hat{\mathbf{P}}_{\text{out}} = f_{\text{nonlinear}}(\hat{\mathbf{P}}) = f_{\text{linear}}(\hat{\mathbf{P}}) + \sum_{i=1}^{E} \mathbf{\gamma}_{i}Z_{i}
\end{equation}

\begin{figure}[ht]
 \centering 
 \includegraphics[width=0.9\columnwidth]{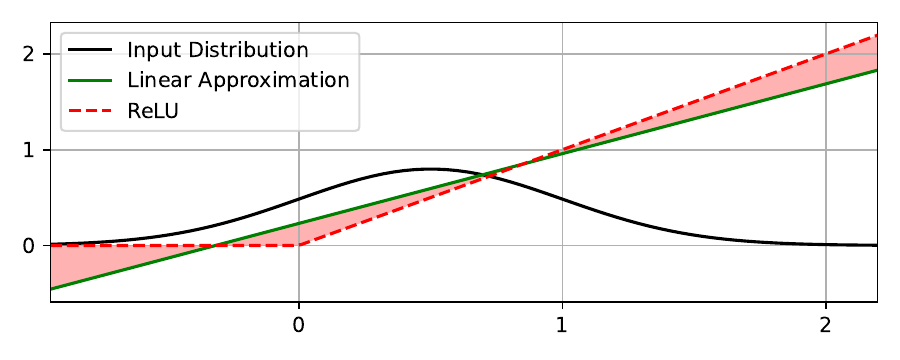}
 \caption{Given the input distribution, we compute a linear approximation of the nonlinear function. The red area represents the approximation error. The actual error is quantified by integrating this red area weighted by the input distribution.}
 \label{fig:nonlinear_diagram}
\end{figure}

The number of extra random vectors $E$ is determined by the input dimensionality of $\hat{\mathbf{P}}$.
Following the approach of Li and Shen \cite{li2024improving}, we employ the least squares method for linear approximation, ensuring the minimization of the mean squared error across the input distribution. \cref{fig:nonlinear_diagram} illustrates this approximation process.
$\hat{\mathbf{P}}_{\text{out}}$ is still an affine form after nonlinear transformation. 
With the affine arithmetic rules, the PAFs can be propagated through a general INR model. 
The number of terms of $Z_i$ 
equals the number of nonlinear functions encountered in the network plus the dimensionality of the input parameters.
This typically large value of $E$ satisfies the conditions for the central limit theorem. 
Therefore, at each network layer, the PAFs converge to Gaussian distributions following this theorem.
Other conditions of the central limit theorem are discussed in detail in the original paper \cite{li2024improving}.

After the uncertainty propagation, we can obtain the output PAF 
\begin{equation} \label{eq:output_PAF}
    \textbf{Y}_{\textbf{x}_0, \hat{\textbf{P}}} = p_{0,\textbf{x}_0} + \sum_{i=1}^{N} p_{i,\textbf{x}_0}Z_{i,\textbf{x}_0}   
\end{equation}
In this equation, $p_{0, \textbf{x}_0}$ and $p_{i, \textbf{x}_0}$ are scalar values because our INR output is one-dimensional.
According to the central limit theorem, this output PAF can be approximated with a Gaussian distribution with $\mu = p_{0, \textbf{x}_0}$ and $\sigma=\sum_{i=1}^{N} p^2_{i, \textbf{x}_0}$. 

\textbf{Covariance Estimation.}
In the above section, we derived the PAF $\textbf{Y}_{\textbf{x}_0, \hat{\textbf{P}}}$, which represents the value distribution across various parameter settings at a specific position $\textbf{x}_0$.
For another position $\textbf{x}_1$, the PAF $\textbf{Y}_{\textbf{x}_1, \hat{\textbf{P}}} = \mathbf{p}_{0,\textbf{x}_1} + \sum_{i=1}^{N} \mathbf{p}_{i,\textbf{x}_1}Z_{i,\textbf{x}_1}$ can also be obtained based on \cref{eq:output_PAF}.
This section will show that the covariance between these two positions can be directly calculated from the PAFs.
For $\textbf{Y}_{\textbf{x}_0, \hat{\textbf{P}}}$ and $\textbf{Y}_{\textbf{x}_1, \hat{\textbf{P}}}$, the first $M$ (dimensionality of the parameter domain) terms of $Z_i$ are the uncertainties introduced by the parameter input, and the rest $N-M$ terms of $Z_i$ are the uncertainties introduced by the INR approximation. According to the bi-linearity property of the covariance, the covariance between these two positions can be calculated as
\begin{equation}
    \text{cov} (\textbf{Y}_{\textbf{x}_0, \hat{\textbf{P}}}, \textbf{Y}_{\textbf{x}_1, \hat{\textbf{P}}}) = \sum_{i=1}^{N} \sum_{j=1}^{N} \mathbf{p}_{i,\textbf{x}_0} \mathbf{p}_{j,\textbf{x}_1} \text{cov}(Z_{i,\mathbf{x}_0}, Z_{j,\mathbf{x}_1}). 
\end{equation}

Except for the uncertainties introduced by the parameter domain input, all other uncertainties are independent, i.e., $\text{cov}(Z_{i,\textbf{x}_0}, Z_{j,\textbf{x}_1}) = 1$ if and only if $i = j \leq M$, otherwise $\text{cov}(Z_{i,\textbf{x}_0}, Z_{j,\textbf{x}_1}) = 0$. Therefore, the above equation can be simplified to
\begin{equation}
    \text{cov} (\textbf{Y}_{\textbf{x}_0, \hat{\textbf{P}}}, \textbf{Y}_{\textbf{x}_1, \hat{\textbf{P}}}) = \sum_{i=1}^{M} \mathbf{p}_{i,\textbf{x}_0} \mathbf{p}_{i,\textbf{x}_1}.
\end{equation}

Having the covariance defined, the Pearson correlation between two positions can be calculated as
\begin{equation}
    \rho_{\textbf{Y}_{\textbf{x}_0, \hat{\textbf{P}}}, \textbf{Y}_{\textbf{x}_1, \hat{\textbf{P}}}} =\frac{\text{cov} (\textbf{Y}_{\textbf{x}_0, \hat{\textbf{P}}}, \textbf{Y}_{\textbf{x}_1, \hat{\textbf{P}}})}{\sigma_{\textbf{x}_0, \hat{\textbf{P}}} \sigma_{\textbf{x}_1, \hat{\textbf{P}}}}.
\end{equation}

In this equation, $\sigma_{\textbf{x}_0, \hat{\textbf{P}}}$ and $\sigma_{\textbf{x}_1, \hat{\textbf{P}}}$ are standard deviations for $\textbf{Y}_{\textbf{x}_0, \hat{\textbf{P}}}$ and $\textbf{Y}_{\textbf{x}_1, \hat{\textbf{P}}}$. All the terms in this equation are known after we propagate the PAFs for $\textbf{x}_0$ and $\textbf{x}_1$.

\subsection{Parameter Exploration}
\label{sect:param_explore}

\textbf{Setup.} In the context of ensemble simulations, a crucial task in parameter domain exploration involves identifying simulation parameters that yield desired outputs. 
These desired outputs are frequently expressed in statistical distributions, offering a comprehensive description of simulation results. 
This task presents two primary challenges: (1) \textit{Estimation of Distribution for a Given Parameter Setting}:
For a spatial sub-domain $\Omega \subset \mathbb{R}^{d}$ and a specific parameter setting $\textbf{P}$, we can estimate the regional distribution using the following equation:
\begin{equation} \label{eq_region_integral2}
	\hat{F}(\Omega, \textbf{P}) = \frac{1}{|\Omega|} \int_{\Omega} F(\textbf{X}, \textbf{P})d\textbf{X},
\end{equation}

where $F(\textbf{X}, \textbf{P})$ represents the simulation output at spatial coordinates $\textbf{X}$ with parameters $\textbf{P}$, and $|\Omega|$ denotes the volume of the sub-domain.

(2) \textit{Parameter Identification for Desired Outputs}:
Given a desired output distribution $\mathcal{D}$, the objective is to find a parameter setting $\textbf{P}_{\mathcal{D}}$ such that $\hat{F}(\Omega, \textbf{P}_{\mathcal{D})}$ closely approximates $\mathcal{D}$. 
Ideally, we aim to find an inverse function:
\begin{equation}
    \hat{F}^{-1}(\mathcal{D}) = (\Omega, \textbf{P}_{\mathcal{D}}),
\end{equation}

However, finding an inverse function is a difficult task.
Therefore, in this section, we will illustrate our approach of employing gradient descent to determine optimal parameter settings and spatial bounds that align with the target output value distribution.

\textbf{Optimization-based Method.} In \cref{sect:spat_explore}, we demonstrated the method for converting parameter or spatial ranges into PAF and its subsequent propagation through the INR.
The output PAF can be interpreted as a Gaussian distribution. 
The calculation is differentiable, enabling defining a loss function between the output Gaussian distribution and a target distribution, and applying backpropagation to optimize the input parameters.
This optimization process can be executed in the spatial domain, the parameter domain, or concurrently in both domains.
To quantify the difference between the output and target distributions, we can employ the Kullback-Leibler (KL) divergence or Jensen–Shannon (JS) divergence as the loss function. From the comparison experiment in the supplemental material, they have similar performance. In the later experiments, we use KL divergence as our loss function.
If the target distribution follows a Gaussian distribution $\mathcal{N}(\mu_\text{target}, \sigma_\text{target})$, the KL divergence can be calculated analytically as 
\begin{equation}
    \label{eq:gaussian_dkl}
    D_{KL}=\ln{\frac{\sigma}{\sigma_\text{target}}} + \frac{\sigma_\text{target} ^2 + (\mu_\text{target} - \mu) ^ 2} {2 \sigma ^ 2} - \frac{1}{2},
\end{equation}

where $\mu$ and $\sigma$ are the mean and standard deviation from the output PAF. Assume the target histogram has $H$ bins with edges $e_1, e_2, ... , e_{H+1}$, where each bin $i$ covers the range $[e_i, e_{i+1})$. The KL divergence between the target histogram and the output Gaussian distribution is calculated as
\begin{equation}
    D_{KL}=\sum^N_{i=1}p_i\ln(\frac{p_i}{q_i}),
\end{equation}

where $p_i$ is the normalized count (probability) of the histogram in the bin $i$, and $q_i$ is the probability mass of the Gaussian distribution in the same bin. 
Instead of integrating the Gaussian PDF numerically from $e_i$ to $e_{i+1}$ to get the exact probability mass $q_i$, we approximate that with the Gaussian probability density at the bin center $c_i = \frac{e_i + e_{i+1}}{2}$ multiplied by the bin width:
\begin{equation}
    q_i = (e_{i+1} - e_{i}) f(c_i | \mu, \sigma),
\end{equation}

where $f(x | \mu, \sigma)$ is the Gaussian PDF. With the loss function, the gradient descent is applied to update the input parameters and positions.

The input range is defined by the minimum coordinate $\textbf{x}_-$ and maximum coordinates $\textbf{x}_+$ of a hypercube in the spatial and parameter domain. 
Each element in $\textbf{x}_-$ should be smaller than the corresponding element in $\textbf{x}_+$ to produce a valid hypercube as the uncertainty propagation input. 
However, gradient descent may find arbitrary values for them. 
To avoid the violation of the above definition, the input range is defined differently during the parameter optimization process to ensure a valid input range. 
Instead of defining a hypercube with $\textbf{x}_-$ and $\textbf{x}_+$, we use its center $\textbf{x}_c$ and the scale $\textbf{x}_s$ along each dimension. 
To make sure the scale is always positive and this cube has a minimum scale, we define it as
\begin{equation} \label{eq_xscale}
    \textbf{x}_s = \textbf{x}_{sqrt} ^ 2 + \beta,
\end{equation}

where $\beta$ is a positive constant for the minimum scale. We do not update the value of $\beta$ in the training and $\textbf{x}_s$ from this equation is always positive. The input minimum and maximum are computed from the center and scale:
\begin{align*}
    \textbf{x}_{min} = & \tanh({\textbf{x}_{c} - \textbf{x}_{s}}), \\
    \textbf{x}_{max} = & \tanh({\textbf{x}_{c} + \textbf{x}_{s}}).
\end{align*}

The $\tanh$ function ensures the minimum and maximum $\textbf{x}$ are in the plausible spatial and parameter range. We can adjust the optimization goal depending on the nature of the target feature to be explored. For example, if the target feature is known to not change in size or position, we can keep $\textbf{x}_s$ or $\textbf{x}_{c}$ as a constant in the optimization.

%% file: tex/4_results.tex
\section{Results}
We evaluated our Explorable INR using cosmology and ocean simulations.
In \cref{sect:exp_setup}, the simulations are explained, and the implementation details of the proposed model are shown.
We then compared our Explorable INR, existing surrogate models, and other INR structures (\cref{sect:inr_eval}).
Due to space limitations, the detailed evaluation of ocean simulations is in the supplementary material.
In \cref{sect:exp_spatial_explore}, we compare the ensemble uncertainty and covariance field obtained from training data, dense sampling, and uncertainty propagation. 
Finally, in \cref{sect:exp_param_explore}, we test the use of gradient descent for uncertainty propagation to efficiently search for the parameter setting corresponding to a given value distribution in the spatial domain.

\subsection{Experiment Setup}
\label{sect:exp_setup}

\textbf{Ensemble Simulations}
Our Explorable INR is evaluated on two ensemble simulation datasets, Nyx \cite{almgren2013nyx} and Model for Predication Across Scales-Ocean (MPAS-Ocean) \cite{ringler2013multi}. The simulations are conducted on a supercomputer with 648 nodes, where each node has an Intel Xeon E5-2680 CPU with 14 cores and 128 GB memory. 28 and 128 processes are used in the Nyx and MPAS-Ocean simulations, respectively.

Nyx is a compressible cosmological hydrodynamics simulation developed by the Lawrence Berkeley National Laboratory. 
The simulation data contains physical features like dark matter density, temperature, and velocity.
In our experiments, we select dark matter density for training and evaluation; thus, the simulation members are regular scalar fields.
A single simulation data size is $512^{3}$ with 32-bit floating points.
The simulation members are sampled from the following parameter domains suggested by scientists, the total matter density $(OmM \in [0.12, 0.155])$, the total density of baryons $(OmB \in [0.0215, 0.0235])$, and the Hubble constant $(h \in [0.55, 0.85])$.
We randomly sampled 130 parameters to run the simulation; 100 for training and 30 for testing.

MPAS-Ocean is a global ocean system simulation developed by Los Alamos National Laboratory. 
The following simulation parameters are suggested by scientists, the amplitude of the ocean surface wind stress $(BwsA \in [0.0, 5.0])$, the critical bulk Richardson number $(CbrN \in [0.25, 1.00])$, the magnitude of the Gent McWilliams mesoscale eddy parameterization $(GM \in [600.0, 1500.0])$, and horizontal viscosity $(HV \in [100.0, 300.0])$.
The simulation was run on 70 random parameter settings from the parameter domain for training and 30 random parameter settings for testing.
The simulation members are unstructured grids, which are comprised of spherical coordinates and ocean temperatures.
The spherical coordinates in the MPAS-Ocean dataset consist of three components: latitude, longitude, and depth, where the depth represents the distance below the ocean's surface.
A single simulation member has 11,845,146 temperature values and the ensemble members share the same spherical coordinates.

\textbf{Model.} 
For the Explorable INR and other baselines, the training and inference in our experiments are done on NVIDIA A100 Tensor GPU 80GB. The Explorable INR was implemented by PyTorch \cite{paszke2019pytorch}.

As explained in \cref{sect:method_inr}, the proposed model uses mixed feature grids for the spatial domain and 1D feature lines for the parameter domain.
The resolution for the 3D grid is $64^{3}$, three 2D planes are $256^{2}$, and $16$ for 1D lines.
The feature vector length is set to 64 for the spatial features and 16 for the parameter features.
The exploration of the impact of feature vector length is demonstrated in the supplementary materials.
The 3D feature grid is initialized by a uniform distribution $U(-10^{-4}, 10^{-4})$, as recommended by Instant-NGP \cite{muller2022instant}. 
The 2D planes are designed to capture small-scale variations and details that might be lost in the low-resolution 3D grid.
Due to the Hadamard Product for fusing the features, initializing the values around $1.0$ allows for the refinement of the spatial features from the 3D grid.
Therefore, we initial the features on the 2D planes using $U(0.999, 1.001)$.
The initialization of the parameter feature lines is $U(0.01, 0.25)$ because this initialization ensures that each parameter contributes equally to the fused parameter feature.
Finally, the feature decoder comprises three layers with 128 hidden nodes each, ReLU activation for hidden layers.

\subsection{Evaluation of Explorable INR}
\label{sect:inr_eval}
We evaluate the Explorable INR from two distinct research perspectives: surrogate model and INR.
The selected state-of-the-art (SOTA) baselines and results are demonstrated in \cref{sect:comp_surro} and \cref{sect:comp_inr}.

\subsubsection{Comparison with Baseline Surrogate models}
\label{sect:comp_surro}
For SOTA surrogate models, we select InsituNet \cite{he2019insitunet}, VDL-Surrogate \cite{shi2022vdl}, and GNN-Surrogate \cite{shi2022GNN}. 
InSituNet is an image-based surrogate model that takes simulation parameters as input and generates corresponding images.
However, it does not support data-level comparison, and a separate InSituNet must be trained for each visual mapping.
On the other hand, VDL-Surrogate is a view-dependent-latent-based surrogate model that can generate simulation raw data through interpolation from different viewpoints, thus enabling data-level comparison metrics such as Peak Signal-to-Noise Ratio (PSNR) and maximum difference (MD).
GNN-Surrogate is another surrogate model which focuses on unstructured grids. They employed an adaptive network to train the surrogate model for simulation outputs of size $10^7$. 
However, the adaptive-resolution strategy is not universally applicable to all datasets.
Thus, we only compare the GNN-Surrogate to our Explorable INR on the MPAS-Ocean dataset.
The qualitative and quantitative comparisons are conducted on the Nyx dataset and discussed in the following paragraphs, with additional evaluations on the MPAS-Ocean dataset presented in the appendix.

\begin{table}[htbp]
\centering
\caption{ The proposed Explorable INR is evaluated on Nyx and MPAS-Ocean datasets. PSNR and MD are selected as data-level metrics for the comparison between our Explorable INR and existing surrogate models: VDL-Surrogate and GNN-Surrogate. }
\label{tab:psnr_md}
\resizebox{\columnwidth}{!}{%
\begin{tabular}{cc|c|c|c}
                            &      & Ours & VDLSurro & GNNSurro \\ \hline
\multirow{2}{*}{Nyx}        & PSNR (dB) & \textbf{45.31}    & 36.38    & N/A      \\
                            & MD   & \textbf{0.1015}   & 0.2235   &          \\ \hline
MPAS-Ocean                  & PSNR (dB) & \textbf{50.81}    & N/A    &  45.54        \\
(global)                    & MD   & \textbf{0.1721}   &    &  0.2528  \\ \hline
MPAS-Ocean                  & PSNR (dB) & \textbf{39.76}    & N/A    &  36.95        \\
(ROI)                       & MD   & \textbf{0.1542}   &    &  0.2106 
\end{tabular}%
}
\end{table}

\textbf{Quantitative Results.}
We benchmark our Explorable INR against VDL-Surrogate and GNN-Surrogate using PSNR and MD, and the results are shown in \cref{tab:psnr_md}.
For the MPAS-Ocean dataset, we examined both global perspectives and specific Regions of Interest (ROI) defined as $160^{\circ}W$ to $80^{\circ}E$, $26^{\circ}S$ to $26^{\circ}N$, and areas up to 200 meters below sea level.
It is important to note that the VDL-Surrogate \cite{shi2022vdl} preprocesses MPAS-Ocean data into a structured grid before training and inference, whereas our Explorable INR is trained directly on the original unstructured grid. Due to this difference, a direct comparison between Explorable INR and VDL-Surrogate on MPAS-Ocean data may not be appropriate.
In contrast, GNN-Surrogate is specifically designed for unstructured grids, making it suitable for a direct comparison with Explorable INR on the MPAS-Ocean dataset.
However, it is not appropriate to compare the Nyx dataset, which has regular grids.

\begin{figure}[htbp]
  \centering 
  \includegraphics[width=\columnwidth]{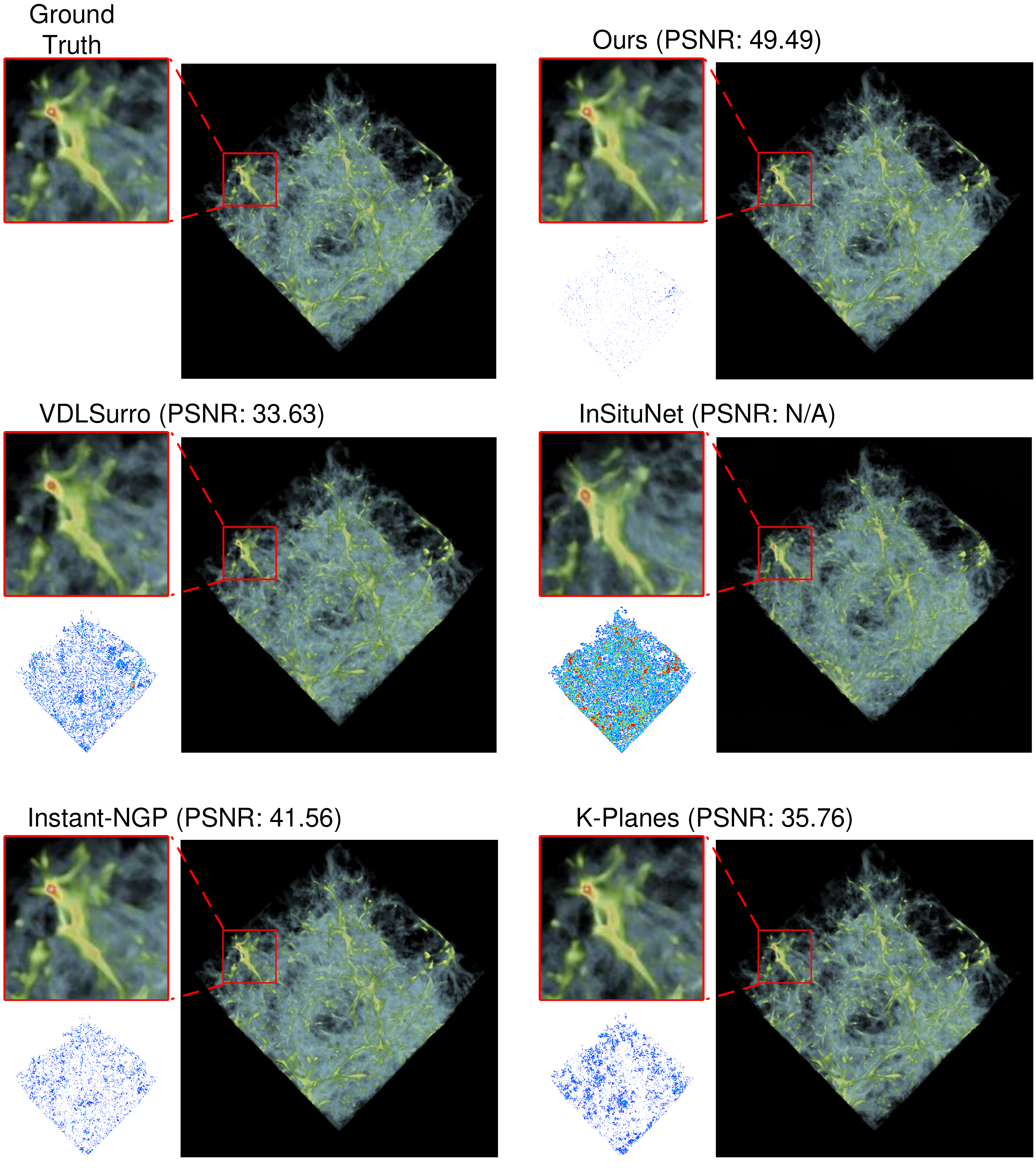}
  \caption{ The comparison of images generated by the Explorable INR, VDL-Surrogate, InSituNet, Instant-NGP, and K-Planes for the Nyx dataset against the ground truth image is presented. The red box highlights the intricate details in the Nyx data.
  The blue/red points stand for the voxel difference between the ground truth and the reconstructed field.
  } 
  \label{fig:qualti_nyx}
\end{figure}

\textbf{Qualitative Results.}
For the images rendered by surrogate models, we will look at the visual fidelity and similarity to the ground truth image.
\cref{fig:qualti_nyx} provides a visual comparison of surrogate models applied to the Nyx dataset, highlighting the performance of the Explorable INR, VDL-Surrogate, and InSituNet.
\cref{fig:qualti_nyx} consists of six sub-images, each representing ground truth or different methods. VDLSurro and InSituNet are existing surrogate models. 
For each method, the sub-image displays the volume rendering results, a zoomed-in patch, and the pixel-level differences between the predicted images and the ground truth.
The red box (zoom-in patch) reveals that the Explorable INR successfully captures critical details such as the white dot within the red dot and a thin line below the red dot, which VDL-Surrogate and InSituNet fail to reconstruct, respectively.
The color coding for pixel-level difference indicates the magnitude of discrepancies: white for low, blue for medium, and red for high differences. 
Based on the pixel-level differences, the Explorable INR demonstrates superior accuracy with the lowest pixel-level differences.

\textbf{Computational and Storage Efficiency}
A fundamental requirement for an effective surrogate model is that its running speed should surpass that of the simulation it represents; otherwise, the utility of employing a surrogate model becomes questionable.
Thus, we compare the execution time and model size for simulations and surrogate models.
\cref{tab:time_size_comp} summarizes the time and model size for the selected methods. 
For simulations, we report the simulation running time and the size of the raw output data.
For surrogate models, we define two key time measures:
(1) Training time: The time for training on 100 simulations for the Nyx dataset and 70 for MPAS-Ocean until convergence.
(2) Inference time: The time for inference of the remaining 30 simulations, for both Nyx and MPAS-Ocean dataset.
\cref{tab:time_size_comp} shows that the execution time of Explorable INR is substantially less than that of simulation and other existing surrogate models. 
Moreover, the size of our model is compact for both Nyx and MPAS-Ocean datasets.
This significant reduction in computational time and data size underscores the efficacy of our proposed method as a surrogate model. 
The Explorable INR not only accelerates the process of obtaining simulation results but also drastically reduces the storage requirements, making it a highly efficient and practical tool for ensemble simulations.

\begin{table}[htbp]
\centering
\caption{
    The execution time for simulation is the simulation running time, while the execution time for surrogate models are training time plus inference time. The size for simulation is the raw data size, and the size for surrogate models is the model size.}
\label{tab:time_size_comp}
\resizebox{\columnwidth}{!}{%
\begin{tabular}{c|cc|cc}
           & \multicolumn{2}{c|}{Nyx}                                 & \multicolumn{2}{c}{MPAS-Ocean}                        \\ \cline{2-5} 
           & \multicolumn{1}{c|}{Execution Time}                    & Size      & \multicolumn{1}{c|}{Execution Time}                 & Size      \\ \hline
Simulation & \multicolumn{1}{c|}{139.8 (hr)}              & 65 GB     & \multicolumn{1}{c|}{82.7 (hr)}            & 903.7 MB  \\ \hline
Ours       & \multicolumn{1}{c|}{18.3 (hr)  + 85.5 (sec)} & 112.17 MB & \multicolumn{1}{c|}{4.6 (hr) + 7.4 (sec)} & 112.17 MB \\ \hline
VDLSurro   & \multicolumn{1}{c|}{52.7 (hr) + 19.3 (sec)}  & 2.01 GB   & \multicolumn{1}{c|}{N/A}                  & N/A       \\ \hline
GNNSurro   & \multicolumn{1}{c|}{N/A}                     & N/A       & \multicolumn{1}{c|}{36 (hr) + 60 (sec)}   & 2.18 GB   
\end{tabular}%
}
\end{table}

\subsubsection{Comparison with Baseline INRs}
\label{sect:comp_inr}
This section commences with an introduction to baseline INRs, also followed by quantitative and qualitative comparisons. 
We have selected Instant-NGP \cite{muller2022instant} and K-Planes \cite{fridovich2023k}, which utilize feature grids in their models, as well as CoordNet \cite{han2022coordnet}, which adapts from SIREN \cite{sitzmann2020implicit} and has been previously applied to scientific data.
Both Instant-NGP and K-Planes encode spatial coordinates in implicit neural representations. 
Instant-NGP employs multi-resolution 3D voxel grids, facilitating rapid optimization and rendering. 
While maintaining comparable performance, K-Planes addresses the memory constraints of Instant-NGP by utilizing 2D planes for spatial encoding. 
CoordNet extends the SIREN architecture \cite{sitzmann2020implicit} by incorporating periodic activation functions and residual blocks, making it particularly suitable for complex scientific data representation.
To compare our hybrid spatial encoding scheme with other INRs, we replace the spatial encoding block of our Explorable INR with either of the state of the art encoding methods Instant-NGP and K-Planes. Then, we train and evaluate the model where the spatial encoding uses that method as opposed to our proposed hybrid encoding.
In addition, we also used comparable numbers of parameters across feature grid methods.
The Explorable INR used a $64^{3}$ cube and three $256^{2}$ planes, while we used $64^{3} + 48^{3} + 44^{3}$ cubes for Instant-NGP and three $391^{2}$ planes for K-Planes.
For CoordNet, we extended its inputs to accommodate simulation parameters, enabling a comparison between feature grid-based INRs and pure multilayer perceptron INRs.
Finally, all of the selected existing INRs are trained for the same number of epochs as training epochs for the Explorable INR.
This selection of baseline INRs allows for a comprehensive evaluation of our Explorable INR against diverse approaches in the field, each with its unique strengths and methodologies for representing and processing complex data.

\begin{table}[htbp]
\centering
\caption{The proposed INR and other INR methods are evaluated Nyx and MPAS-Ocean datasets. PSNR and MD are selected as the data-level metrics for the comparison between the Explorable INR and other INR methods: Instant-NGP, c, and CoordNet.}
\label{tab:compare_inr}
\resizebox{\columnwidth}{!}{%
\begin{tabular}{c|cc|cc}
           & \multicolumn{2}{c|}{Nyx}                              & \multicolumn{2}{c}{MPAS-Ocean}                        \\
           & \multicolumn{1}{c|}{PSNR (dB)}      & MD              & \multicolumn{1}{c|}{PSNR (dB)}      & MD              \\ \hline
Ours       & \multicolumn{1}{c|}{\textbf{45.31}}          & \textbf{0.1015} & \multicolumn{1}{c|}{\textbf{50.81}} & \textbf{0.1721} \\ \hline
Instant-NGP & \multicolumn{1}{c|}{44.89} & 0.1122          & \multicolumn{1}{c|}{49.62}          & 0.1889          \\ \hline
K-Planes    & \multicolumn{1}{c|}{32.91}          & 0.2115          & \multicolumn{1}{c|}{45.29}          & 0.2001          \\ \hline
CoordNet   & \multicolumn{1}{c|}{14.96}          & 0.3982          & \multicolumn{1}{c|}{10.78}          & 0.5066         
\end{tabular}%
}
\end{table}

\textbf{Quantitative Results.}
We compared the performance of the Explorable INR and other existing INRs using PSNR and MD metrics, as in our comparisons with other surrogate models. 
The results, presented in \cref{tab:compare_inr}, demonstrate that our method outperforms other INR approaches.
Among the feature grid-based methods, our method performs better than Instant-NGP and K-Planes, proving our claim in \cref{sect:method_inr}.
For CoordNet, it has the worst performance among all INRs.
Two plausible explanations can be proposed for this observation:
First, CoordNet's architecture uses fully connected layers for spatial and parameter inputs. This approach, lacking separate components for processing these inputs, may hinder learning the hierarchical nature of ensemble simulations and capturing the broad influence of simulation parameters on the entire field.
Second, the model might require a significantly longer training duration and a greater number of epochs to adequately learn and represent the intricacies of the complex dataset.

\textbf{Qualitative Results.}
Similar to our comparison with existing surrogate models, we evaluate the visual fidelity and similarity to the ground truth image for the INR-based approaches. 
Note that due to space constraints and CoordNet's inferior performance, we have omitted its rendering results from the visual comparison. The rendered image of CoordNet is a uniform green cube, indicating its failure to effectively learn the ensemble dataset.
The voxel-wise difference analysis reveals that the spatial encoding employed in our Explorable INR outperforms those of Instant-NGP and K-Planes.
This visual assessment, coupled with the quantitative metrics, underscores the efficacy of our Explorable INR's spatial encoding mechanism. It highlights our model's capacity to capture and represent complex datasets more accurately than other SOTA INR approaches, even when they are integrated into our pipeline framework.

\subsection{Spatial Exploration Experiment}
\label{sect:exp_spatial_explore}
In this section, we apply the proposed approach from \cref{sect:spat_explore} to obtain the physical feature distribution for a specific spatial location across ensemble members.
Specifically, we aim to compute the mean and variance values for each location in the field.
Furthermore, we also compute the covariance between a given reference location and all the other locations in the field.

\begin{figure}[t]
  \centering 
  \includegraphics[width=\columnwidth]{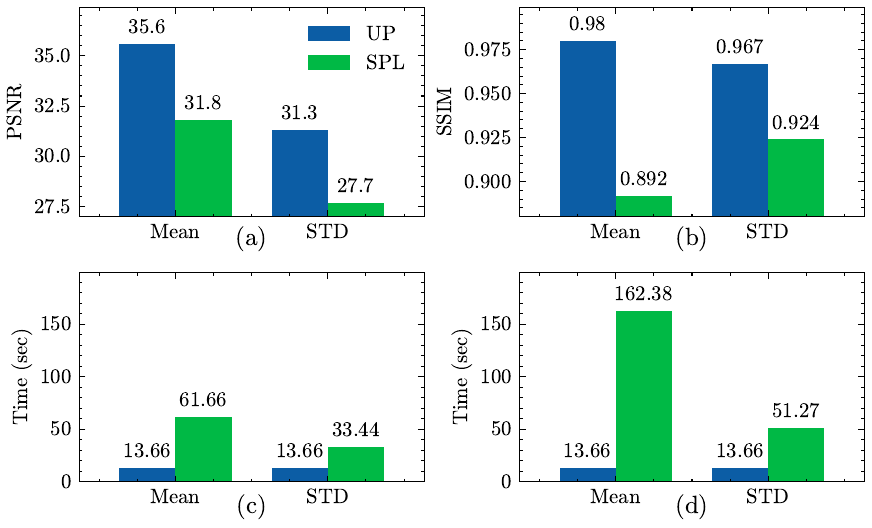}
  \caption{%
    The figure compares the performance of uncertainty propagation (UP) and sampling (SPL) methods through statistics on Nyx results.
    Given the same running time, (a) and (b) are the PSNR and similarity index measures (SSIM) of mean and standard deviation (STD) for both methods.
    Given the same PSNR and SSIM of mean and STD, (c) and (d) are the running time for both methods to achieve the value.
  } 
  \label{fig:nyx_barchart}
\end{figure}

\subsubsection{Ensemble Uncertainty}
Scientists use the Nyx simulation to study dark matter density under various conditions. They want to understand how density values change on average. Here, we compare three methods by the mean density values and standard deviations.
The first method is simulation data (SIM). We use the 130 simulations to calculate this mean and STD. This value may not reflect the true density distribution because of the limited number of simulation runs. However, we use them as the ground truth since better statistics cannot be obtained unless we run more simulations.
The second method is sampling the Explorable INR (SPL). It is much more efficient to query the model than to run the simulation. Therefore, it is feasible to query more samples in the parameter domain in an acceptable time, such as a few minutes, and obtain the density mean and STD. 
The third method is the proposed uncertainty propagation (UP) method. It reduces the dense sample cost when inferring the Explorable INR model. Since we make approximations in our approach to speed up the calculation, UP takes much less computation time but might be less accurate than SPL compared to the ground truth (SIM).

\begin{figure}[htbp]
  \centering 
  \includegraphics[width=\columnwidth]{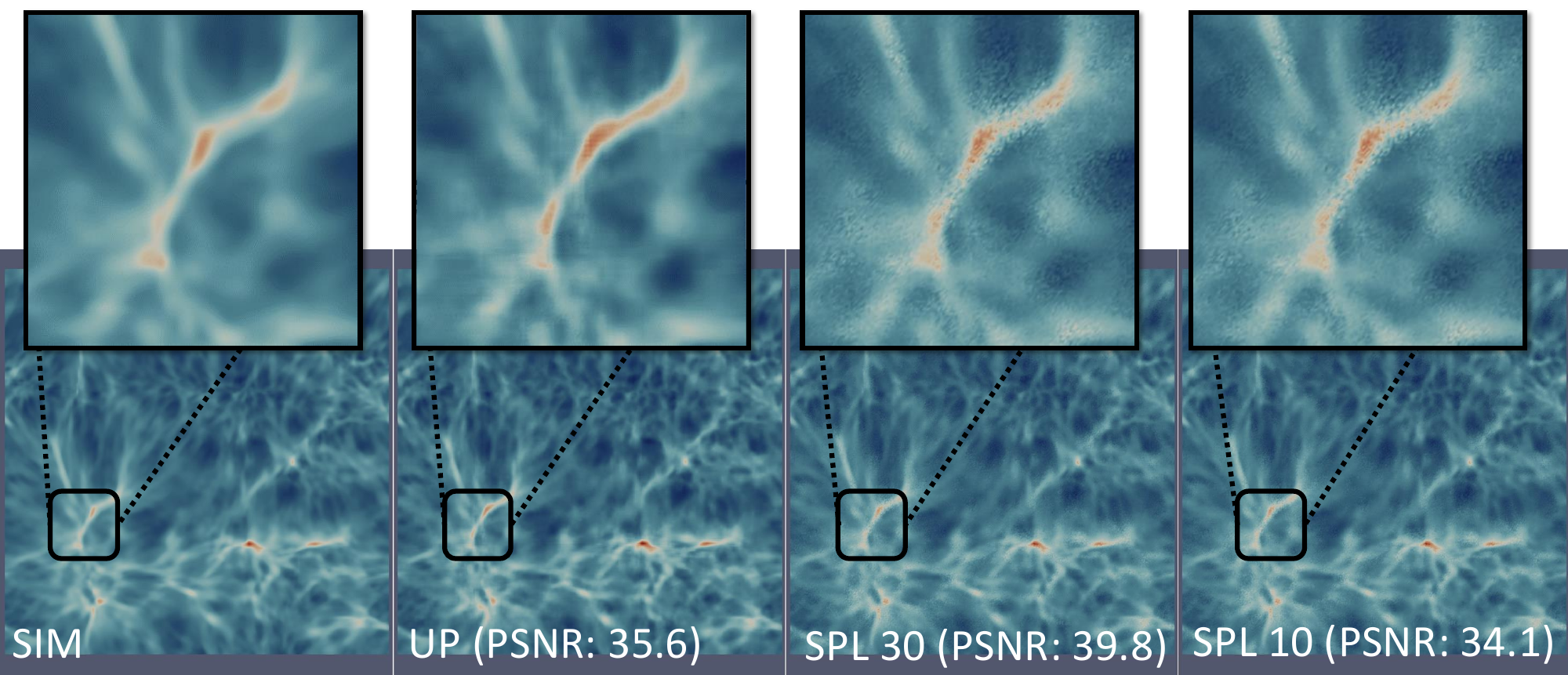}
  \caption{%
    The volume rendering images of the mean value field. SIM stands for the simulation output. UP stands for uncertainty propagation. ``SPL 30'' and ``SPL 10'' stand for the Monte Carlo method with 30 and 10 samples. We highlight the features of interest in the image.
  } 
  \label{fig:mean_qualitative}
\end{figure}

To evaluate the performance of UP, we compare UP and SPL from two perspectives: accuracy and running time. 
For the UP method, the running time and the quantitative metrics for mean and STD are fixed.
For the SPL method, sampling more parameter settings will give more accurate results, so the quantitative metrics of mean and STD increase as the sampling number increases.
\cref{fig:nyx_barchart} (a) and (b) show the accuracy of both methods in PSNR and SSIM. Since the UP has a fixed running time (13.66 sec), we run the SPL method using the same computation time.
The results show that UP performs better than SPL for both PSNR and SSIM of mean and STD.
\cref{fig:nyx_barchart} (c) and (d) show the running time of both methods for given PSNR and SSIM. 
The results show that to achieve the same performance as UP, SPL takes about 2-13 times of computation time.

We show the rendering results of the mean fields across ensemble members using four methods in \cref{fig:mean_qualitative}. All methods can clearly show the high-density regions of interest. However, UP is more similar to the ground truth SIM than the sampling methods. The sampling method tends to underestimate the high densities and introduce noises especially when the number of samples is small. We show these noises in the zoomed-in image in \cref{fig:mean_qualitative}. UP produces a smooth mean value field without noise. In summary, although the computation time and the quality matrix do not significantly differ between UP and SPL with less than 30 samples, UP is free from random sample noises and produces smoother mean value fields.

\begin{figure}[htbp]
  \centering 
  \includegraphics[width=\columnwidth]{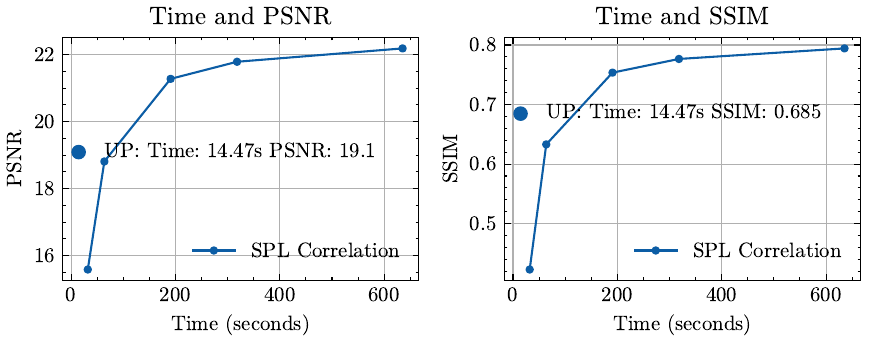}
  \caption{
    UP approach is a point due to the fixed running time and performance, while SPL's running time and performance depend on the number of samples.
    Left: the relationship between the sampling time and PSNR. Right: the relationship between the sampling time and SSIM. 
  } 
  \label{fig:corr}
\end{figure}

\subsubsection{Pearson Correlation}
Spatial correlation is another important metric to domain scientists. Similar to the ensemble uncertainty, we compute the covariance field using SPL, SIM, and UP. For the correlation field, we pick a reference spatial point with maximum mean density. The correlation field is the Pearson correlation between all points and the reference point. The time and accuracy relation is in \cref{fig:corr}. We can see a similar trend compared to the mean and variance fields. Compared to SPL, UP is more accurate when the computation time is the same and faster when the accuracy is the same.
Though the PSNR from UP and SPL are relatively low compared to the correlation field from the simulation outputs, they are similar to each other with PSNR 28.79. This indicates that our calculation for the correlation field accurately represents the correlation learned by the Explorable INR, which can be slightly different from the correlation calculated from the available simulation runs. Due to limited simulation data, correlation calculations between spatial point pairs may not accurately represent true relationships. The correlation field generated by the INR can serve as an exploratory tool, guiding domain scientists to investigate regions of interest through targeted simulations. To understand the scientists' perspective on our method, an interview with a scientist is included in the supplemental material.

\begin{figure}[t]
  \centering 
  \includegraphics[width=\columnwidth]{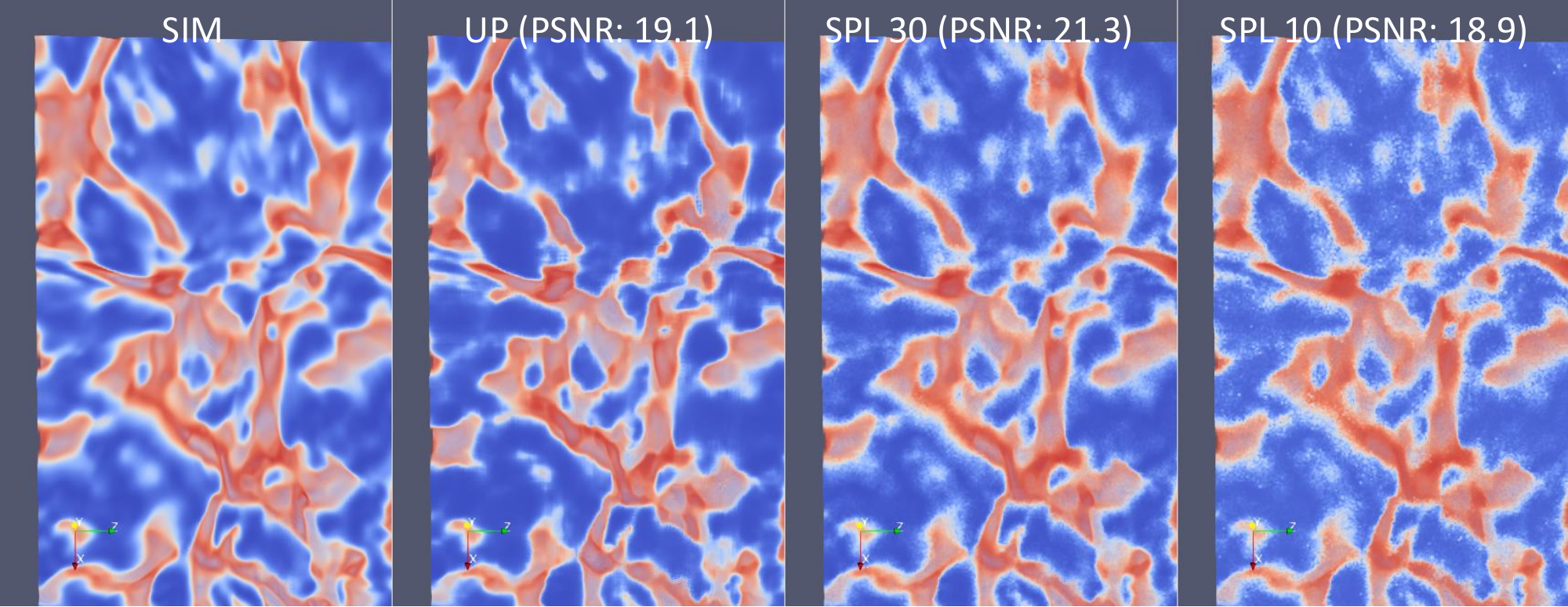}
  \caption{%
    The volume rendering results for the Pearson correlation field. SIM stands for the simulation output. UP stands for uncertainty propagation. ``SPL 30'' and ``SPL 10'' stand for the Monte Carlo method with 30 and 10 samples. Four images are rendered with the same transfer function.
  } 
  \label{fig:corr_qualitative}
\end{figure}

The volume rendering results of the correlation field are shown in \cref{fig:corr_qualitative}. We can see the spatial correlation structures are similar across different methods. Similar to the mean value field in \cref{fig:mean_qualitative}, we observe the random sample noises in the sampling method when the number of samples is small.

\subsection{Parameter Exploration Experiment}
\label{sect:exp_param_explore}
In this section, we apply the proposed approach in \cref{sect:param_explore} to the Nyx dataset for identifying the location and simulation parameters for the desired output distribution. 
We first briefly explain how the experiment is conducted. 
A target feature of interest is selected by examining several visualization results from the Explorable INR model. 
Since the scalar values of this feature conform well to a Gaussian distribution, we calculate its mean and standard deviation. In other situations, the target distribution can be determined by the prior knowledge or by building a histogram of the selected feature of interest. In the Nyx simulation, assume the feature size does not change across the simulation parameters, so we keep the $\mathbf{x}_s$ in \cref{eq_xscale} as constant and optimize the parameters and the feature centers by \cref{eq:gaussian_dkl}. The gradient descent is run for 1000 iterations and kept the parameters when the KL divergence was less than $10^{-5}$. Every time we find a feature candidate, we multiply the learning rate by $10$ in the next optimization step to move out of the local minimum. 

\begin{figure}[htbp]
  \centering 
  \includegraphics[width=\columnwidth]{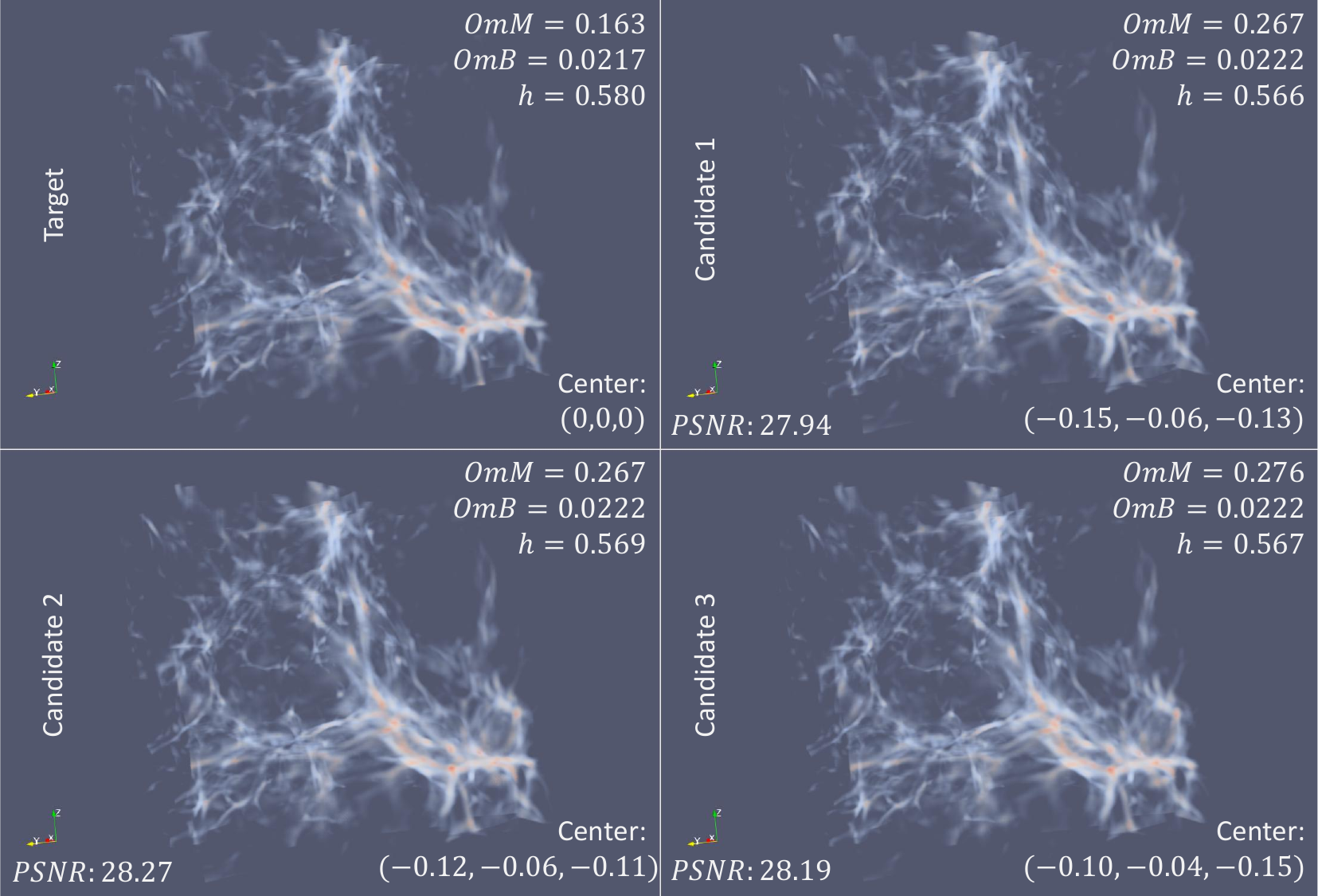}
  \caption{%
    The target feature and the found candidates in our parameter optimization experiments. The parameters and the region centers are noted. PSNRs between candidates and the target are also shown. The overall structure is very similar between the target and the candidates. 
  } 
  \label{fig:parameter_optimization}
\end{figure}

The target feature and three found candidates are in \cref{fig:parameter_optimization}. The overall structure is very similar between the target and the candidates, with only small differences in the details. The parameters of the candidates are not close to the target, showing that they are similar features found in different parameter settings. 
In terms of the optimization time, our experiments take $5.6$ seconds to perform 1000 iterations and find $16$ candidates. This optimization time is negligible compared to the manual examination of the data.

%% file: tex/5_conclusion.tex
\section{Discussion and Future works}

In this section, we discuss the benefits of the proposed approach compared with existing surrogate models, the limitations of this work, and our future studies.

The proposed Explorable INR has the following benefits, related to the limitations of some or all existing methods.
\begin{enumerate}
    \item Arbitrary resolution of output data.
    \item Faster offline training speed and smaller model size.
    \item Flexibility for exploring different visual mappings.
    \item Efficiency for ensemble members analyses.
\end{enumerate}
First, the Explorable INR allows querying arbitrary coordinates within a defined spatial domain, enabling data and image generation at any resolution. This surpasses the resolution limitations of existing models like InSituNet, VDL-Surrogate, and GNN-Surrogate. The flexibility comes from incorporating INR, which provides greater freedom for spatial domain input, unlike the constrained output resolutions of other models due to their network structures.
Second, the Explorable INR offers superior training efficiency and model size compared to alternatives. As discussed in the appendix, its training time is significantly faster than that of existing works, and the model size is much smaller than that of those works. These improvements stem from using a feature grid-based model, which updates fewer parameters per training data point, and a novel design combining feature grids and planes for optimized efficiency and compactness.
Third, the Explorable INR, like GNN-Surrogate and VDL-Surrogate, overcomes InSituNet's limitations in handling large joint spaces of simulation parameters and visual mappings by predicting raw data instead of images.
Finally, unlike existing surrogate models that require dense sampling in the parameter domain for ensemble analysis, the Explorable INR allows direct calculation of statistical summaries at arbitrary spatial positions through uncertainty propagation, enhancing ensemble member analysis capabilities.

One limitation of our model is that visual artifacts appear in the mean value field due to the non-smooth patterns on the feature grid cell boundary. This happens when feature values change sharply between neighboring cells. Our current setup uses a $64^{3}$ cube and $256^{2}$ planes, which can create noticeable artifacts in high-quality images. To fix this, future work could explore different data structures or add smoothing techniques to create more natural transitions between cells.
Another limitation of our approach is its assumption that normally distributed data, while prevalent in scalar field data due to underlying physical principles in scientific simulations, may not always hold true. The PAF may exhibit less accuracy when the target data deviates from a Gaussian distribution. Future research directions could explore implementing Gaussian mixture models to better approximate diverse probability distributions in the target data.

\section{Conclusion}
In this work, we introduce the Explorable INR, a novel INR-based surrogate model, that allows for querying of values at specific locations and parameter settings, significantly reducing memory and computational costs compared to existing models. The proposed model is based on the feature grid approach, which plays a crucial role in enhancing the efficiency and accuracy of exploring ensemble simulations. 
Our Explorable INR employs four strategies, including dividing the spatial and parameter domains into separate feature grids, applying the Hadamard product to fuse spatial and parameter feature vectors, using the mix of a feature grid and feature planes for the spatial encoding, and utilizing 1D feature lines for the parameter encoding, to effectively address the challenges of high computational costs and memory requirements.
In addition, we can efficiently explore the spatial and parameter spaces in ensemble simulations through uncertainty propagation. For spatial space exploration, we demonstrate the method for efficiently obtaining statistical distributions of values across input regions. For parameter space exploration, the developed method enables efficient identification of physical attribute distributions and suitable parameters that match desired distributions.